\documentclass{article}

\usepackage{arxiv}

\usepackage[utf8]{inputenc} 
\usepackage[T1]{fontenc}    
\usepackage{hyperref}       
\usepackage{url}            
\usepackage{booktabs}       
\usepackage{amsfonts}       
\usepackage{nicefrac}       
\usepackage{microtype} 
\usepackage{amsmath}
\usepackage{mathtools}
\usepackage{amssymb}
\usepackage{subcaption}

\usepackage{lipsum}		
\usepackage{graphicx}
\usepackage{natbib}
\usepackage{doi}
\usepackage{amsthm}
\usepackage{xcolor}

\newtheorem{thm}{Theorem}[section]  
\newtheorem{lemma}[thm]{Lemma}  
\newtheorem{corollary}[thm]{Corollary}
\newtheorem{definition}[thm]{Definition}

\newtheorem{remark}[thm]{Remark}
\newtheorem{assumption}[thm]{Assumption}

\title{The Disparate Effects of Partial Information in Bayesian Strategic Learning}

\author{%
  Srikanth Avasarala \\
 Georgia Institute of Technology \\
  \texttt{savasarala9@gatech.edu} \\
  \And
  Serena Wang \\
  Harvard University \\
  \texttt{ serenalwang@g.harvard.edu} \\
  \And
  Juba Ziani \\
  Georgia Institute of Technology \\
  \texttt{jziani3@gatech.edu} \\
}

\date{}

\renewcommand{\shorttitle}{\textit{arXiv} Template}

\hypersetup{
pdftitle={The Disparate Effects of Partial Information in Bayesian Strategic Learning},
pdfsubject={q-bio.NC, q-bio.QM},
pdfauthor={Srikanth Avasarala, Juba Ziani},
pdfkeywords={First keyword, Second keyword, More},
}

\begin{document}
\maketitle

\begin{abstract}
We study how partial information about scoring rules affects fairness in strategic learning settings. In strategic learning, a learner deploys a scoring rule, and agents respond strategically by modifying their features---at some cost--—to improve their outcomes. However, in our work, agents do not observe the scoring rule directly; instead, they receive a \emph{noisy signal} of said rule. We consider two different agent models: (i) \emph{naive} agents, who take the noisy signal at face value, and (ii) \emph{Bayesian} agents, who update a prior belief based on the signal.

Our goal is to understand how disparities in outcomes arise between groups that differ in their costs of feature modification, and how these disparities vary with the \emph{level of transparency} of the learner's rule. For naive agents, we show that utility disparities can grow unboundedly with noise, and that the group with lower costs can, perhaps counter-intuitively, be disproportionately harmed under limited transparency. In contrast, for Bayesian agents, disparities remain bounded. We provide a full characterization of disparities across groups as a function of the level of transparency and show that they can vary \emph{non-monotonically} with noise; in particular, disparities are often minimized at \emph{intermediate} levels of transparency. Finally, we extend our analysis to settings where groups differ not only in cost, but also in prior beliefs, and study how this asymmetry influences fairness.
\end{abstract}.


\section{Introduction}
Machine learning systems are increasingly being deployed in high-stakes domains such as college admissions, lending, and hiring. A common assumption in the design of these models is that the data encountered at test time is drawn from the same distribution as the training data. However, this assumption breaks down when individuals \emph{strategically} adapt their features in response to the deployed model. In real life, individuals may invest in test preparation, tune their resumes to match job descriptions and automated CV reviewing algorithms, or open dummy credit accounts to artificially improve metrics like credit usage---all in an effort to improve their predicted outcomes. This phenomenon, known as \emph{strategic learning}, was first formalized by~\citet{hardt2016strategic}, and has become central to understanding robustness algorithmic decision-making.

While such strategic behavior may be rational from an individual perspective, it can have undesirable implications for society and in particular for fairness. When individuals differ in the cost of modifying their features---due to resource constraints, institutional barriers, or historical disadvantage---strategic adaptation can reinforce or amplify existing inequalities. Prior work has shown that disparities in the cost of feature manipulation can translate into disparities in both scoring and classification outcomes even when the underlying model treats all individuals equally~\citep{milli2019social, hu2019disparate}.

A common, key assumption in much of the strategic classification literature is that individuals have \emph{full knowledge} of the deployed model and can best respond accordingly. Yet this assumption is often unrealistic. In practice, scoring rules are rarely transparent. Banks and credit bureaus do not disclose the exact algorithms behind loan decisions or credit scores; recidivism prediction tools such as COMPAS operate as proprietary black boxes~\citep{angwin2016machine}; and AI-based hiring platforms use complex and often non-interpretable models to rank candidates. Therefore, addressing fairness in strategic learning environments also requires thinking about how individuals perceive deployed rules and algorithms, both when all agents have access to the same information about the deployed rule and when there are informational disparities across individuals.

These observations raise a critical question: how does \emph{partial knowledge} of the model affect the fairness of strategic learning systems? More precisely, how do different levels of transparency---ranging from full disclosure to total opacity---influence disparities in outcomes across populations? And how do these effects depend on the assumptions we make about how individuals operate under uncertainty? In this work, we develop a framework to study fairness under partial information in strategic learning environments. We model agents who observe only a \emph{noisy signal} of the deployed decision rule. We consider two types of agent responses to this signal: (i) \emph{naive} agents, who treat the signal as ground truth and optimize against it directly~\citep{jagadeesan2021alternative}; and (ii) \emph{Bayesian} agents, who combine the signal with a prior belief to form a posterior, and optimize based on this posterior belief on the deployed rule~\citep{cohen2024bayesian}. Our analysis focuses on how outcome disparities—--measured both in terms of model scores and individual utilities (that also take costs into account)—--arise across groups that differ in their cost of feature modification.  We examine how these disparities evolve as a function of the amount of information the learner releases about the model. 

\paragraph{Summary of Contributions.} Our paper makes the following contributions:
\begin{itemize}
    \item In Section~\ref{sec:naive-eq}, we analyze the fairness of \textbf{naive agents} under partial transparency. We show that while \emph{score} disparities remain constant, \emph{utility} disparities vary monotonically with the level of noise in the model signal. Surprisingly, the group with lower costs of feature change can be disproportionately harmed by noisy information, leading to unbounded disparities in utility, due to ``over-spending'' on ineffective modifications. 
    \item In Section~\ref{subsec:fairbay}, we turn to \textbf{Bayesian agents} and fully characterize how score and utility disparities depend on the prior and noise level of the released signal. We characterize when disparities arise as a function of the parameter of the problem. Perhaps surprisingly, we show that disparities can be \emph{non-monotone} in the level of information revealed by the learner and are minimized at intermediate transparency levels. 
    \item In Section~\ref{sec:fair-uneq}, we characterize disparities when groups differ not only in cost but also in their \textbf{prior beliefs}, expanding the model of~\citet{bechavod2022information}. We derive bounds on group disparities as a function of the \emph{information overlap} of~\citet{bechavod2022information}. 
\end{itemize}

\paragraph{Related work}

This work lies at the intersection of fairness and strategic behavior in algorithmic decision-making. A growing body of research has studied how individuals adapt their features to secure better outcomes from predictive models---known as \emph{Strategic Classification} or \emph{Strategic Learning}---, and how such behavior leads to disparate impacts across populations.

Strategic Learning was initially introduced by \citet{hardt2016strategic} and sparked a large area of research studying how agents respond to decision rules in learning systems~\citep{braverman2020role,dong2018strategic,zhang2022fairness,lechner2023strategic,chen2020learning,ahmadi2021strategic,sundaram2023pac} to only name a few. For a recent survey, please refer to~\citep{podimata2025incentive}. 

While early work primarily focused on agents ``gaming'' classifiers---in the context of loans, this could be seen as opening dummy credit cards to artificially lower credit utilization and inflate credit scores---, more recent papers have considered actual improvements---e.g., improving one's ability to repay loans on time---, as seen in the works of~\citet{kleinberg2020classifiers,shavit2020causal,harris2021stateful,bechavod2022information,ebrahimi2024double}.

A significantly smaller subset of this literature explicitly tackles fairness. Notably, \citet{milli2019social} and \citet{hu2019disparate} show that unequal ability to manipulate features (in the form in unequal feature manipulation costs) can lead to unfair outcomes. Other works examine disparities arising from population-level variation in feature distributions or strategic behavior itself \citep{jung2020fair,liu2020disparate}. Most recently,~\citet{liu2025fairness} propose a dynamic pricing algorithm that ensures fairness by constraining strategic group misreporting.

Our work builds directly on recent advances in modeling strategic classification under partial information. The noisy response model introduced by~\citet{jagadeesan2021alternative} assumes agents do not see the exact classifier---rather, they see a noisy version of the deployed classifier, and ``naively'' best respond to this noisy signal. 
We build on the model of~\citet{jagadeesan2021alternative} by considering the disparate impacts of information revelation when such naive agents have different costs of modifying their features. Furthermore, we additionally consider a Bayesian model of agent behavior, where agents form and update beliefs about the decision rule based on observed information---rather than taking a noisy and potentially inaccurate signal at face value. This approach directly follows the recent work of~\citet{cohen2024bayesian}, who first introduce Bayesian agents in the context of strategic classification. Our main departure from the work of~\citet{cohen2024bayesian} is that we study Bayesian agents strategic classification from a \emph{fairness} perspective, and aim to understand how the information revealed to agents affects group-level disparities.

Our model is perhaps most closely related to \citet{bechavod2022information}, who also consider agents without access to the true model and study resulting group outcomes. However, we note two key differences:
\begin{itemize}
\item \emph{(i) Modeling of partial information:} they assume agents infer the model from peer samples, whereas we allow agents to observe noisy outputs of the model directly. Further, agents in \citet{bechavod2022information} are not Bayesian, and instead compute the deterministic ``most reasonable'' model based on the observed samples;
\item \emph{(ii) Role of the learner:} in our framework, the learner can control how much information is revealed via a tunable noise parameter, whereas in their work the learner has no direct influence over agent beliefs.
\end{itemize}

Finally, our approach connects with recent work on strategic classification with causal and informational structure \citep{ahmadi2022classification, horowitz2023causal, efthymiou2025incentivizing}. While \citet{efthymiou2025incentivizing} study optimal effort allocation under uncertain causal graphs and classifiers, their focus is primarily on incentivizing agents to modify features in ``desirable'' ways. Our work differs by directly analyzing how information policies shape fairness under incomplete information.

\section{Model and preliminaries}

We study a Stackelberg game between a \emph{learner} (or principal) and a population of \emph{agents} where each agent belongs to one of $m$ sub-populations (or "groups"). We focus on the case of $m = 2$ in this work\footnote{Groups are statistically independent of each other, so all insights for $m = 2$ can be generalized to any $m$}. 

Let the sub-populations (or groups) be denoted $g_1$ and $g_2$. Each agent $i$ has a feature vector $x_i \in \mathbb{R}^d$, and is assigned a label $y_i \in \mathbb{R}$ by the learner. We assume\footnote{As is common in related work---see for example~ \citep{bechavod2022information}.} that the learner assigns scores \emph{linearly}, i.e. that their score is given by
\[
y_i = x_i^\top \theta_*,
\]
where $\theta_* \in \mathbb{R}^d$ is the 
rule deployed by the learner.

However, each agent observes only partial information about the deployed model $\theta_*$, in the form of a noisy signal about said model. Formally, and as in \citep{jagadeesan2021alternative}, each agent observes a signal $S$ of the model $\theta_*$ corrupted by an additive noise, given by:
\begin{equation}
    S = \theta_* + \sigma Z.
\end{equation}
where $\sigma \in \mathbb{R}_{\ge 0}$ is the parameter controlling the amount of variance in the noise, and $Z$ is an independent sample from a standard Gaussian---i.e., we assume that $Z \sim \mathcal{N}(\mathbf{0},\mathbf{I}_d)$. This signal can be interpreted as the information that the \emph{learner} decided to reveal about their classifier; the lower the value of $\sigma$, the more information they reveal to agents. For example, credit scoring companies reveal \emph{partial} information about their models (i.e., the weight they put on each feature, or which features they consider, but not how these features are computed). Our goal is to understand \emph{how the level of information revelation by the learner affects disparities across populations} under varying i) costs and ii) initial information about the learner's model. 
\paragraph{Principal-Agent interaction}
Agents first observe the signal $S$ then strategically change their features to try to improve their score. For a given agent, we denote $x$ their original features and $x'$ their modified features. We also write 
$
\Delta x \triangleq x' - x
$
for simplicity of notation. To do so, they first build a posterior belief on $\theta_*$ based on their signal $S$, denoted by $\theta \sim \pi^S$. Given modified feature vector $x' \in \mathbb{R}^d$, an agent's utility function for moving from $x$ to $x'$ is given by
\begin{equation}
    u(x,x'; g) = \text{score}(x';g) - c(x,x';g),
\end{equation}
where
\[
\text{score}(x';g) = \mathbb{E}_{\theta \sim \pi_g^S}\left[x^{\prime \top} \theta\right]
\]
is the expected score the agent gets under posterior $\pi_g^S$, and 
\[
c(x,x';g) = \frac{1}{2}(x' - x)^\top A_g (x' - x)
\]
is a Mahalanobis-distance-based cost function\footnote{Such $\ell_2$ cost functions are common in the strategic learning literature, e.g.~\citep{bechavod2022information,cohen2023sequential}} for changing agent features from $x$ to $x'$. The cost function is parametrized by $A_g \in \mathbb{R}^{d \times d}$, which is called the cost matrix for group $g$ and is assumed to be positive definite (PD)\footnote{Similar assumptions are made in related work, e.g.~\citet{bechavod2022information}}. 

We now consider two principled ways that agents interpret the signal $S$: (i) a \emph{naive} agent that takes the signal $S$ at face value, i.e., has the posterior belief $\hat{\theta} = S$, as in~\citep{jagadeesan2021alternative}; (ii) a \emph{Bayesian} agent that computes their \emph{posterior belief} on $\hat{\theta}$ based on both a prior belief $\pi$ about the model and the signal $S$; this is the model of~\citet{cohen2024bayesian}. Formally: 
\begin{itemize}
\item (i) \emph{Naive agent:} 
In the naive case, the agent solves the following optimization problem:
\begin{align}\label{eq:opt-naive}
     \hat{x}_g &= \arg\max_{x'}~~~ u(x,x'; g) \nonumber \,\\
    &= \arg\max_{x'}~~~ S^\top x' - \frac{1}{2}(x' - x)^\top A_g (x' - x). \nonumber \,\\
\end{align}
\item (ii) \emph{Bayesian agent:} In this case, an agent in group $g$ has a prior distributional belief $\pi$ on the learner’s
model, encoding their initial knowledge of the model. 
After
observing realization $s$ of signal $S$, the agent updates their prior in a Bayesian fashion to a new
posterior belief given by $\pi_g^S (\theta) = \mathbb{P}_{\theta \sim \pi} \big[\hat{\theta} = \theta \vert S \big]$.  This posterior is a distributional belief on
the learner’s model, to be interpreted as the perceived chance that the model is $\theta$. The individual
then changes their features from $x$ to $x'$ to maximize the utility conditioned on the observed signal in the following way
\begin{align} \label{eq:opt-bay}
 \hat{x}_g &= \arg\max_{x'}~~~ \mathbb{E}_{\theta \sim \pi_g^S}~~ u(x,x' ; g \vert S) \nonumber \,\\
    &= \arg\max_{x'}~~~ \mathbb{E}_{\theta \sim \pi_g^S } \left[{\theta}^\top x' | S \right] - \frac{1}{2}(x' - x)^\top A_g (x' - x).
\end{align}
For the purpose of our analysis, we consider the same amount of information revealed across both groups. For a \emph{Bayesian} agent, we assume the agent prior in a group to be $\pi_g \sim \mathcal{N}(\omega_g, \gamma^2 \mathbf{I}_d)$, where $\gamma^2$ is the variance of each component of the prior distribution vector. 
\end{itemize}


\paragraph{Fairness metrics} The main goal of the paper is to understand disparities across groups when it comes to strategically fitting the deployed decision rule. To do so, we now define two different measures of disparity that we use to quantify fairness between two groups in equilibrium, i.e. after agents in both the groups change their features. 
\begin{enumerate}
\item\emph{Score disparity: }The \emph{score disparity} $\mathcal{F}_s$ 
is defined as 
\begin{align}
\mathcal{F}_s 
&= \mathbb{E} [\Delta~ \text{score}_1]
- \mathbb{E} [\Delta~ \text{score}_2] \nonumber\,\\
& = \mathbb{E} [{\theta_*}^\top  \Delta x_1]
- \mathbb{E}[{\theta_*}^\top  \Delta x_2].
\end{align}

\item \emph{utility disparity: } The utility disparity $\mathcal{F}_u$ 
is defined as
\begin{align}
\mathcal{F}_u &= \mathbb{E}[\Delta u_1] - \mathbb{E}[\Delta u_2]   \nonumber \,\\
 &= \mathbb{E} [{\theta_*}^\top \Delta x_1 - \Delta x_1^\top A_1 \Delta x_1]\nonumber \,
 \\&-\mathbb{E} [{\theta_*}^\top  \Delta x_2 - \Delta x_2^\top A_2 \Delta x_2].
\end{align}
\end{enumerate}
The first definition, \emph{score disparity}, measures the difference in average score improvement between the two groups, capturing how changes in features translate to perceived gains across populations~\citep{bechavod2022information}. The second definition, \emph{utility disparity}, extends this notion by incorporating the cost of change, reflecting the net benefit each group experiences after accounting for adjustment effort. 


We now define regions of score and utility disparity based on their signs with respect to the parameterization $\mathcal{P}$.
\begin{definition}
We define the following three regions corresponding to $\mathcal{P}$ when using the metric $\mathcal{F}$
 \begin{enumerate}
     \item \emph{Exploitation:} A parameterization corresponds to \emph{Exploitation} if the metric $\mathcal{F}> 0$.
     \item \emph{Neutrality:} A parameterization corresponds to \emph{Neutrality} if the metric $\mathcal{F}= 0$. 
     \item \emph{Burden:} A parameterization corresponds to \emph{Burden} if the metric $\mathcal{F}< 0$. 
 \end{enumerate}   
\end{definition}

The first definition, \emph{Exploitation}, implies that the more advantaged group has higher improvement for the choice of parameters. \emph{Neutrality} implies equal improvements shown by both groups. 
Finally, \emph{Burden} indicates the region where the more advantaged group suffers a lower improvement from strategic change of feature.

\paragraph{Cost disparities} The associated cost matrices for groups $1$ and $2$ are denoted $A_1$ and $A_2$ (respectively) and parameterized by $\mathcal{A} :=(A_1, A_2)$. Importantly, in the rest of the paper, we assume that there are cost disparities between groups $1$ and $2$---similarly to the works of~\citet{hu2019disparate} and~\citet{milli2019social}---, and in particular, that group $1$ is \emph{advantaged} and incurs systematically \emph{lower cost} compared to group $2$. Formally: 

\begin{assumption}
$A_1 \prec A_2$, or equivalently, $A_2-A_1$ is positive-definite.
\end{assumption}
\section{Preliminaries: Agent Best Responses}

In this section, we compute the feature improvement vector $\Delta x_g$ for an agent in a population $g$ for both (i) \emph{naive} agents and (ii) \emph{Bayesian} agents by solving their optimization problems---respectively given by Eq~\ref{eq:opt-naive} and Eq~\ref{eq:opt-bay}. Proofs of both lemmas are provided in Section~\ref{ap:eq} of the Appendix.


 \begin{lemma}\label{le:delxface}
For a Naive agent in group $g$, the feature change $\Delta x_g$ in equilibrium is given by:
\begin{equation}
    \Delta x_g  = A_g^{-1}S_g.
\end{equation}
\end{lemma}

 \begin{lemma}\label{le:delxbay}
For a Bayesian agent in group $g$, the feature change $\Delta x_g$ in equilibrium is given by:
\begin{equation}
    \Delta x_g= A_g^{-1}  [\omega_g + \beta(\sigma, \gamma)(S_g - \omega_g)]
\end{equation}
where $ \beta(\sigma, \gamma) := \frac{\gamma^2}{\gamma^2 + \sigma^2}$, and $S_g = \theta_* + \sigma Z_g$.
\end{lemma}

For the Naive agent, observe that the feature change does not depend on the noise level $\sigma$. Intuitively, this is because the Naive agent takes the signal at face value and does not try to compensate for potential randomness in the signal. In contrast, the Bayesian agent's response depends on a function of the noise of the signal, $\sigma$, and the noise of the prior, $\gamma$.

\section{Fairness under Incomplete Information}\label{sec:fair-analysis}

Using the obtained analytical forms for improvements, we now perform fairness analyses using the defined metrics $\mathcal{F}_s
$ and $\mathcal{F}_u$ for both a \textit{naive} and \textit{Bayesian} agent. Relevant proofs for this section can be found in Appendix~\ref{ap:fairness}.

\subsection{\textit{Naive} agent}\label{sec:naive-eq}

In this section, we characterize how disparities in outcomes arise as a function of the amount of information revealed by the learner for \emph{naive} agents. 

\paragraph{\emph{Score disparity}} We start by providing a characterization of the \emph{score disparity}, and show that it is \emph{invariant} to the amount of information $\sigma$ released by the learner:

\begin{thm}\label{le:scf_naive}
    For a \textit{naive} agent, in equilibrium:
    \begin{equation} \label{eq:es_naive}
        \mathcal{F}_s = {\theta_*}^\top (A_1^{-1}  - A_2^{-1}) {\theta_*}.
    \end{equation}
\end{thm}

\begin{corollary}\label{co:scf_naive} The \emph{score disparity} satisfies $\mathcal{F}_s  > 0$ and is invariant w.r.t $\sigma$.
\end{corollary}

\begin{corollary}\label{co:var-sc}
    $\text{Var}( {\theta_*}^\top  \Delta x_1 -  {\theta_*}^\top \Delta x_2)$
 is monotonically increasing in $\sigma$.
\end{corollary}

For a \emph{naive} agent, the \emph{score disparity} is only related to the cost matrix $A_g$ and the true model ${\theta_*}$. We note that this score is always positive, i.e. it always shows \textit{Exploitation}. While  \emph{\emph{score disparity}} of a \textit{naive} agent is agnostic
to signal variance or information, it is important to note that the differences in variances of scores in both groups is monotonic in $\sigma$. This reflects variability in score improvement across individuals in a group, which gets nullified when computing the average score-improvement across the group.

\paragraph{\emph{Utility disparity}} On the other hand, the \emph{utility disparity} shows a monotonic trend from favoring the advantaged group at higher information to unboundedly favoring the disadvantaged group at the limit of no information. 
\begin{thm}\label{le:utf-naive}
    For a naive agent, in equilibrium the \emph{utility disparity}  $\mathcal{F}_u$ is given by:
    \begin{equation}\label{eq:ut-naive}
       \mathcal{F}_u =  \frac{1}{2}\big[{\theta_*}^\top (A_1^{-1}  - A_2^{-1}) {\theta_*} - \sigma^2 (\text{Tr}(A_1^{-1}) - \text{Tr}(A_2^{-1}))\big],
    \end{equation}
    where \textit{Tr}(.) denotes matrix trace.
\end{thm}

\begin{figure}
    \centering
    \includegraphics[width=0.5\textwidth]{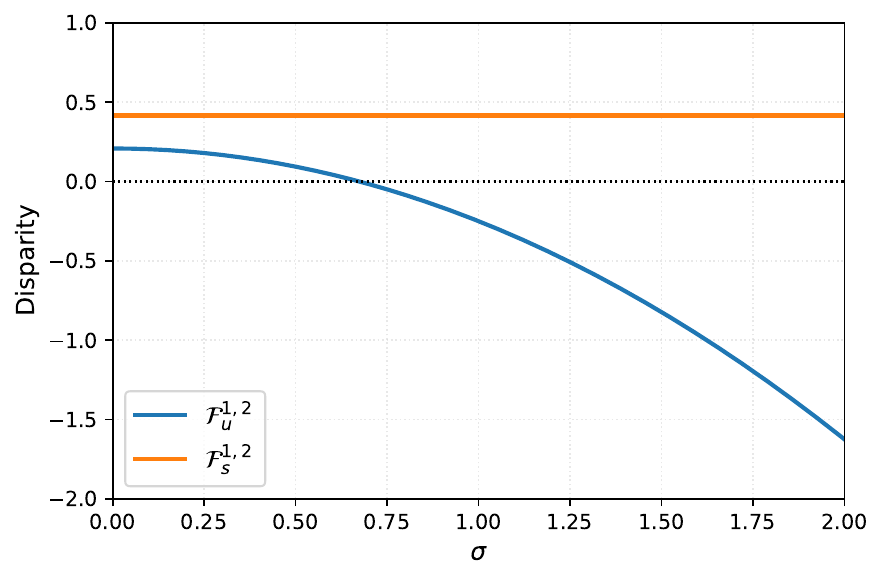}
    \caption{Comparison of \emph{score disparity} $\mathcal{F}_s$ and \emph{utility disparity} $\mathcal{F}_u$ as a function of $\sigma$ for a \textit{naive} agent, computed from Lemma.~\ref{le:scf_naive} and Lemma.~\ref{le:utf-naive} respectively. The dotted black line at zero of y-axis represents \textit{Neutrality}.  The parameters chosen are: $\theta_* = [1,~ 0.5]^\top, A_1 = diag(2,1), A_2 = diag(4,3)$}
    \label{fig:disparity-naive}
\end{figure}
\begin{lemma}\label{co:naiv-ut-beh}
    For a naive agent in equilibrium the utility disparity $\mathcal{F}_u$ is  monotonically decreasing as a function of $\sigma$.  \textit{Neutrality} occurs at $\sigma_{r} = \sqrt{\frac{{\theta_*}^\top (A_1^{-1}  - A_2^{-1}) {\theta_*}}{\text{Tr}(A_1^{-1}) - \text{Tr}(A_2^{-1})}}$. \textit{Exploitation} occurs for $\sigma < \sigma_r$ and \textit{Burden} occurs for $\sigma > \sigma_r$.  Moreover, $\lim_{\sigma \to 0^+} \mathcal{F}_u(\sigma) = \frac{\mathcal{F}_s(\sigma)}{2}$ and $\lim_{\sigma \to \infty} \mathcal{F}_u = -\infty$.
\end{lemma}

The behavior of a \textit{naive} agent is intuitive: as the noise level $\sigma$ increases, agents observe the model at face value, without accounting for the underlying uncertainty. This leads them to rely on distorted signals, resulting in misallocated effort---often overspending on modifying features that have little impact on their score. Consequently, the lower-cost group, which can afford greater strategic feature changes, may in fact expand more cost to modify features due to overconfidence without increasing their score accordingly. Mathematically, as $\sigma$ increases, the increased cost disparity given by the term $\sigma^2(Tr(A_1^{-1}) - Tr(A_2^{-1})) > 0$ in Eq~\ref{eq:ut-naive} (see Proof.~\ref{pr:naiv-ut}) dominates the disparities in score, indicating that the lower cost group is, counter-intuitively, overspending compared to the high-cost group. 


\begin{figure*}
    \centering
    \includegraphics[width=\textwidth]{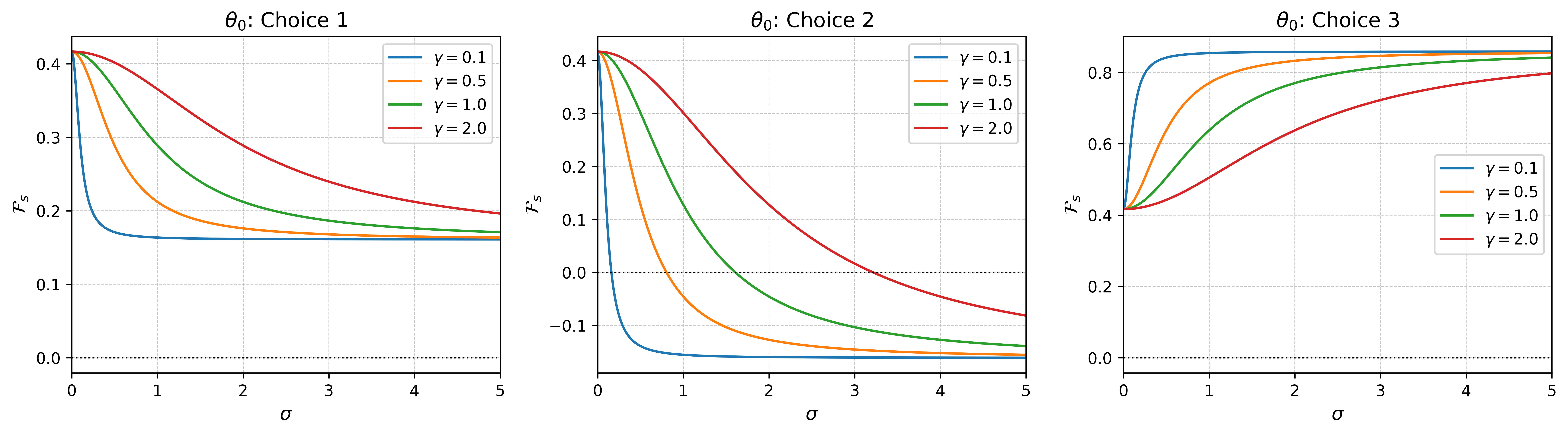}

    \begin{subfigure}[t]{0.32\textwidth}
        \phantomsubcaption
        \label{fig:disparity-bayesian-score-a}
    \end{subfigure}
    \begin{subfigure}[t]{0.32\textwidth}
        \phantomsubcaption
        \label{fig:disparity-bayesian-score-b}
    \end{subfigure}
    \begin{subfigure}[t]{0.32\textwidth}
        \phantomsubcaption
        \label{fig:disparity-bayesian-score-c}
    \end{subfigure}

    \caption{\emph{Score disparity} $\mathcal{F}_s$ as a function of $\sigma$ for a \textit{Bayesian} agent, across different values of $\gamma$. Each panel corresponds to a different prior mean $\theta_0$: 
    \textbf{(a)} $\theta_0 = [0.5,~2]^\top$, 
    \textbf{(b)} $\theta_0 = -[0.5,~2]^\top$, 
    \textbf{(c)} $\theta_0 = k[0.5,~2]^\top$ with $\|\theta_0\| = 2\|\theta^*\|$. 
    The dotted black horizontal line indicates \textit{Neutrality}. Parameters: $\theta^* = [1,~0.5]^\top$, $A_1 = \operatorname{diag}(2,1)$, $A_2 = \operatorname{diag}(4,3)$.}
    \label{fig:disparity-bayesian-score}
\end{figure*}

Relevant plots for both kinds of disparities for a \textit{naive} agent are shown in Fig.~\ref{fig:disparity-naive}. Interestingly, here, partial information revelation is optimal in terms of fairness: it is in the best interest of the learner interested in fairness to reveal $\sigma_r$ amount of information, which achieves the minimum possible disparities in utilities while not affecting score disparities. This may however make it difficult for a policy maker in practice, over simpler mechanisms that just reveal all or no information about the deployed rule. This also implies a fairness-utility trade-off that complexifies a policy maker or learner's decision making: it is easy to see that utility for each of the group is maximized under full information, but fairness requires the learner to hold some information back.

\subsection{Bayesian Agent}
\label{subsec:fairbay}
In this section, we compute the disparities between groups for Bayesian agents at equilibrium. We assume the same prior parameters for agents of both groups throughout and denote $\theta_0 \triangleq \omega_1 = \omega_2$ as the common prior mean, and $\gamma$ the standard deviation of each component of the characteristic vector. We do so to isolate the effect of information parameter $\sigma$ and the cost disparities on fairness, that arise \emph{even when all agents have access to the same information}. 

In this section, we show that the \emph{score disparity} is bounded and monotonic. We then show that \textit{Neutrality} occurs at limited information
if and only if the prior mean is `poorly-aligned' in some sense with the true model. On the other hand, we show that the \emph{utility disparity} shows a transition from being monotonic to non-monotonic in the \emph{level of information} $\sigma$ revealed by the learner; this transition happens at a given, computable value of the \emph{prior}'s standard deviation $\gamma$. We show conditions where \textit{Neutrality} occurs at incomplete information, along with a natural condition where \textit{Neutrality} occurs at several distinct points (in terms of signal variance $\sigma^2$) at incomplete information. 

\paragraph{\emph{Score disparity}} We start by providing a full characterization of score disparities. Before doing so, we remind the reader that $A_1 \prec A_2$, implying in particular that $A_1^{-1} - A_2^{-1}$ is a symmetric positive-definite matrix. For a given matrix $A$, we denote by $\sqrt{A}$ the unique positive-definite matrix $Q$ such that $Q^\top Q = A$. For notational convenience, we define the following transformed variables:
\[
k_{\theta_0} := \sqrt{A_1^{-1} - A_2^{-1}}\, \theta_0 \quad \text{and} \quad k_{\theta_*} := \sqrt{A_1^{-1} - A_2^{-1}}\, \theta_*,
\]

\begin{thm}\label{le:sc-bay}
    For a Bayesian agent, in equilibrium:
    \begin{align}\label{eq:sc-bay-eq}
        \mathcal{F}_s 
        = k_{\theta_0}^\top k_{\theta_*} + ( k_{\theta_*}^\top k_{\theta_*} -k_{\theta_0}^\top k_{\theta_*} ) \beta_{\gamma}(\sigma)
        = (1 - \beta_{\gamma}(\sigma)) k_{\theta_0}^\top k_{\theta_*} + \beta_{\gamma}(\sigma) k_{\theta_*}^\top 
 k_{\theta_*},
    \end{align}
    where $\beta_{\gamma}(\sigma) := \beta(\sigma, \gamma) = \frac{\gamma^2}{\gamma^2 + \sigma^2}$.
\end{thm}
From Eq~\ref{eq:sc-bay-eq}, we observe that $\mathcal{F}_s$ is a weighted sum of two terms: $k_{\theta_0}^\top k_{\theta_*}$, weighted by a factor proportional to $\sigma^2$, and $k_{\theta_0}^\top k_{\theta_*}$, weighted by a factor proportional to $\gamma^2$. Both terms quantify alignment within the subspace spanned by the dominant eigenvectors of the matrix $A_1^{-1} - A_2^{-1}$. Specifically, the first term captures the alignment between the prior mean $\theta_0$ and the ground truth $\theta_*$, while the second term measures the concentration of the ground truth within this subspace.
We now characterize all possible behaviors for $\mathcal{F}_s$:
\begin{lemma}\label{co:sc-bay-prop}
In equilibrium, $\mathcal{F}_s$ is bounded, and is monotonic in $\sigma$. Specifically: 
\begin{enumerate}
    \item if $k_{\theta_0}^\top k_{\theta_*} < k_{\theta_*}^\top k_{\theta_*}$, then $\mathcal{F}_s$ is decreasing in $\sigma$.
    \item if $k_{\theta_0}^\top k_{\theta_*} > k_{\theta_*}^\top k_{\theta_*}$ ($k_{\theta_0}^\top k_{\theta_*} \geq k_{\theta_*}^\top k_{\theta_*}$) , then $\mathcal{F}_s$ is increasing (non-decreasing) in $\sigma$.
    \item At boundaries, $\mathcal{F}_s(0) = k_{\theta_*}^\top k_{\theta_*}$ and $\lim_{\sigma \to \infty} \mathcal{F}_s(\sigma) = k_{\theta_0}^\top k_{\theta_*}$.

\end{enumerate}
\end{lemma} 
The monotonic behaviors of \emph{score disparity} as a function of $\sigma$ are illustrated in Figure~\ref{fig:disparity-bayesian-score-a} (decreasing), Figure~\ref{fig:disparity-bayesian-score-b} (decreasing), and Figure~\ref{fig:disparity-bayesian-score-c} (increasing). Essentially, the conditions compare i) the alignment between the prior mean and the ground truth to ii) the concentration of the ground truth. Intuitively, when ii) is dominant, the initial prior matters less, and information revelation is important to an agent. The more information is revealed (i.e. the smaller the $\sigma$), the more the advantaged group can leverage their cost edge over the disadvantaged group---$\mathcal{F}_s$ is bigger at smaller $\sigma$.

The preceding lemma outlines the general monotonic behavior of $\mathcal{F}_s$ based on the alignment between the prior and true models. In the Corollary below, we provide a condition where Neutrality exists for $\mathcal{F}_s$.
\begin{lemma}\label{co:bay-sc-neut}
  In equilibrium, $\mathcal{F}_s$ shows \emph{Neutrality} if and only if $k_{\theta_0}^\top k_{\theta_*} < 0$, at $\sigma_r = \sqrt{- \frac{k_{\theta_*}^\top k_{\theta_*}}{k_{\theta_0}^\top k_{\theta_*}}} \gamma$.
\end{lemma}

The optimal fairness-oriented revelation policy for the learner, with respect to \emph{score disparity} depends on the behavior of $\mathcal{F}_s$. For instance, if $\mathcal{F}_s$ is increasing, then the best policy is to reveal full information—since $\mathcal{F}_s$ attains its minimum at $\sigma = 0$, though this may be difficult to implement in practice. On the other hand, if $\mathcal{F}_s$ is decreasing, the optimal strategy is to reveal partial or no information, depending on whether \emph{Neutrality} occurs (as shown in Lemma~\ref{co:bay-sc-neut}); in such cases, revealing partial information around $\sigma_r$ may be ideal. Lemma~\ref{co:bay-sc-neut} shows that one must also account for the extent of deterioration caused by the negative term $k_{\theta_0}^\top k_{\theta_*}$, which is weighted higher as $\sigma$ increases. If \emph{Neutrality} does not occur, then revealing no signal is the optimal policy.

We now state a natural condition that is particularly useful for the learner's analysis.
 \begin{corollary}\label{co:assum}
      If $\Vert k_{\theta_0} \Vert \leq \Vert k_{\theta_*} \Vert$, then, in equilibrium:
     \begin{enumerate}
    \item $\mathcal{F}_s$ is monotonically non-increasing w.r.t $\sigma$.
    \item $\mathcal{F}_s$ is bounded by $(k_{\theta_0}^\top k_{\theta_*},k_{\theta_*}^\top k_{\theta_*}]$.
\end{enumerate}
 \end{corollary}   
Note that we use $\Vert.\Vert$ to denote the Euclidean norm for vectors and the spectral norm (i.e., largest singular value) for matrices.  The assumption on the prior mean  $\theta_0$  in Corollary~\ref{co:assum} ensures that the norm of $\theta_0$ is comparable to that of the ground truth $\theta_*$. More precisely, if  $A_g = \alpha_g \mathbf{I}$, then the condition simplifies to  $\Vert \theta_0 \Vert\leq \Vert \theta_* \Vert$. This ensures that the prior does not assign disproportionately more ``magnitude'' or regularization than the true signal, maintaining consistency with the expected scale of the ground truth\footnote{We believe this to be a relatively mild assumption. This occurs, for example, if a user perfectly knows the top $k$ features, but ignores all other features, due to bounded rationality.}.

This also highlights that monotonic increase in $\mathcal{F}_s$ is expected to be less common, typically arising only in unconstrained settings. However, there do exist cases where such increasing behavior still occurs as shown in Figure~\ref{fig:disparity-bayesian-score-c}.

\paragraph{\emph{Utility disparity}}
We now characterize the \emph{utility disparity} for a Bayesian agent in equilibrium. We adopt the same notation as in the \emph{score disparity} analysis, specifically defining the following constants: $k_{\theta_0} = \sqrt{A_1^{-1} - A_2^{-1}}~\theta_0$ and $k_{\theta_*} = \sqrt{A_1^{-1} - A_2^{-1}}~\theta_*$.
\begin{thm}\label{le:bay-ut-eq}
Let 
$m \triangleq (\text{Tr}(A_1^{-1}) - \text{Tr}(A_2^{-1}))$, the \emph{utility disparity} for a Bayesian agent in equilibrium is given by:
\begin{align} \label{eq:bay-ut}
    \mathcal{F}_u(\sigma) &= -\frac{\beta_\gamma^2 (\sigma)}{2} \left(\big( k_{\theta_*} -  k_{\theta_0} \big)^2 + \sigma^2 m\right)  +\beta_\gamma (\sigma)\big( k_{\theta_*} -  k_{\theta_0} \big)^2+ \Big( k_{\theta_0}^\top k_{\theta_*} - \frac{k_{\theta_0}^\top k_{\theta_0}}{2} \Big), 
\end{align}
where $\beta_{\gamma}(\sigma) := \beta(\sigma, \gamma) := \frac{\gamma^2}{\gamma^2 + \sigma^2}$.
\end{thm}

The above expression leads to different cases in terms of both monotonicity and whether we observe \emph{Neutrality}, \emph{Exploitation}, or \emph{Burden}. These cases are highlighted below:

 \begin{lemma}\label{le:bay-ut-cases}
In equilibrium, $\mathcal{F}_u$ for a Bayesian agent, is bounded. Additionally, let $\gamma_c \coloneq \sqrt{\frac{2}{m}} ~ \vert k_{\theta_*} - k_{\theta_0}\vert$, the following are possible behaviors for $\mathcal{F}_u$:

\noindent \textbf{Monotone case:} If $\gamma \leq \gamma_c$, then $\mathcal{F}_u$ is monotonically decreasing in $\sigma$. The following are sub-cases:
\begin{enumerate}
    \item If $k_{\theta_0}^\top k_{\theta_0} > 2 k_{\theta_0}^\top k_{\theta_*}$, then $\mathcal{F}_u$ attains \textit{Neutrality} at a unique $\sigma$.
    \item If $k_{\theta_0}^\top k_{\theta_0} \leq 2 k_{\theta_0}^\top k_{\theta_*}$, then $\mathcal{F}_u$ shows \textit{Exploitation} throughout for all $\sigma$.
\end{enumerate}

\noindent \textbf{Non-monotone case:} If $\gamma > \gamma_c$, then $\mathcal{F}_u$ is non-monotone and attains a global minimum at $\sigma_{\min}$ given by 
    \begin{align}\label{eq:sigma_min}
        \sigma_{min} &= \sqrt{\frac{1}{1 - \frac{\gamma_c^2}{ \gamma^2}}}~ \gamma.
    \end{align}
$\mathcal{F}_u$ is decreasing for $\sigma \leq \sigma_{\min}$, and increasing for $\sigma > \sigma_{min}$. The following are sub-cases:
\begin{enumerate}
    \item If $k_{\theta_0}^\top k_{\theta_0} > 2 k_{\theta_0}^\top k_{\theta_*}$, then $\mathcal{F}_u$ attains \textit{Neutrality} at one point for some $\sigma < \sigma_{\min}$.
    \item If $k_{\theta_0}^\top k_{\theta_0} \leq 2 k_{\theta_0}^\top k_{\theta_*}$, then $\mathcal{F}_u$ has: no point of Neutrality if $\mathcal{F}_u (\sigma_{\min}) > 0$, one if $\mathcal{F}_u (\sigma_{\min}) = 0$, two if $\mathcal{F}_u (\sigma_{\min}) < 0$.
\end{enumerate}

\begin{figure*}
    \centering
    \includegraphics[width=0.81\textwidth]{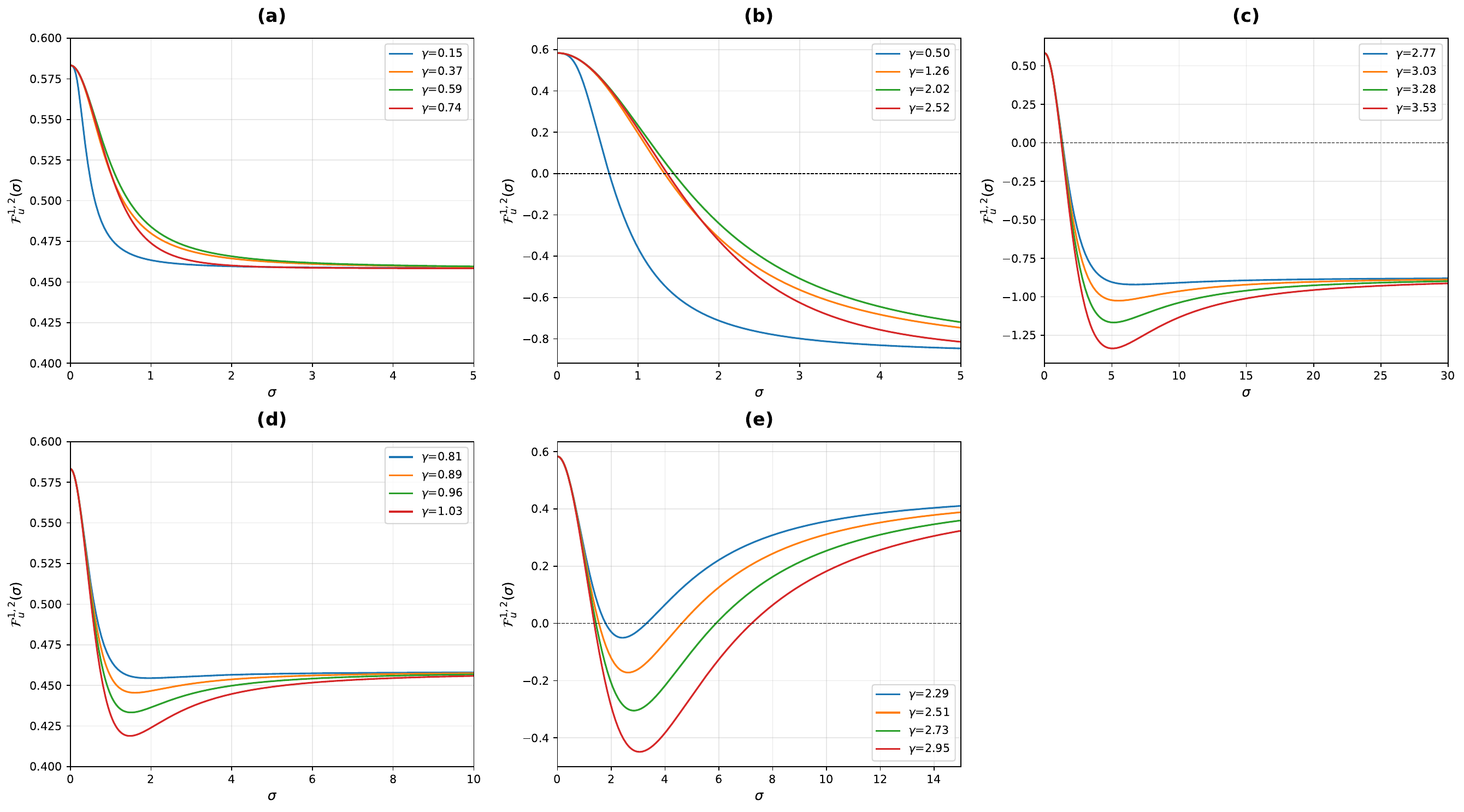}
    
    \begin{minipage}[t]{0.8\textwidth}
        \phantomsubcaption\label{fig:panel-a}
        \phantomsubcaption\label{fig:panel-b}
        \phantomsubcaption\label{fig:panel-c}
        \phantomsubcaption\label{fig:panel-d}
        \phantomsubcaption\label{fig:panel-e}
    \end{minipage}
    \caption{\emph{Utility disparity} $\mathcal{F}_u$ as a function of $\sigma$ for a \textit{Bayesian} agent, for different values of $\gamma$, where each panel—\textbf{(a)} through \textbf{(e)}—represents a chosen prior mean $\theta_0$. The dotted black line on the y-axis denotes \textit{Neutrality}.}
    \label{fig:disparity-bayesian-util}
\end{figure*}

 \end{lemma}

Further, from Eq~\ref{eq:bay-ut} it is easy to see the following values of $\mathcal{F}_u$ in boundary cases:
\[ 
\mathcal{F}_u(0) = \frac{k_{\theta_*}^\top k_{\theta_*}}{2} , ~~~\text{and}~~\lim_{\sigma \to \infty} \mathcal{F}_u (\sigma) = k_{\theta_0}^\top k_{\theta_*} - \frac{k_{\theta_0}^\top k_{\theta_0}}{2}.
\]
Figure~\ref{fig:disparity-bayesian-util} illustrates separate panels, each corresponding to one of the cases described in Lemma~\ref{le:bay-ut-cases} for \emph{utility disparity}. Panel~\ref{fig:panel-a} shows monotonic-decreasing pattern with only \emph{Exploitation}. In contrast, Panel~\ref{fig:panel-b} shows the same monotonic behavior with a single point of \emph{Neutrality}. Panels~\ref{fig:panel-c} through~\ref{fig:panel-e} display non-monotonic behaviors: Panel~\ref{fig:panel-c} features one point of \emph{Neutrality}, Panel~\ref{fig:panel-d} shows none, and Panel~\ref{fig:panel-e} presents two distinct \emph{Neutrality} points.

We observe that for lower prior variance (i.e., $\gamma \leq \gamma_c$), $\mathcal{F}_u$ is monotonically decreasing. In contrast, for higher variance (i.e., $\gamma > \gamma_c$), $\mathcal{F}_u$ becomes non-monotonic, exhibiting a unique global minimum at $\sigma_{\min}$. In this regime, $\mathcal{F}_u$ decreases for $\sigma \leq \sigma_{\min}$ and increases for $\sigma > \sigma_{\min}$.
In the monotonic case, the behavior---whether it is purely decreasing with no \emph{Neutrality} or decreasing with a single \emph{Neutrality} point---is closely aligned with the scenarios analyzed in score disparities (Case 1 of Lemma~\ref{co:sc-bay-prop}). Consequently, the learner’s optimal revelation strategy for \emph{utility disparity} mirrors that for \emph{score disparity} in these cases—either revealing no information or revealing partial information around the point of \emph{Neutrality}, depending on whether it occurs.

An interesting observation is that in both score and \emph{utility disparity}, the condition $k_{\theta_0}^\top k_{\theta_*} < 0$ guarantees similar behavior—i.e monotonic shift from \emph{Exploitation}$\rightarrow$ \emph{Neutrality} $\rightarrow$ \emph{Burden} as $\sigma$ increases. This is intuitive, as the condition implies that the prior is misaligned with the deployed model, in which case information revelation provides an edge to the advantaged group (since the advantaged group can take advantage of knowing the model better than the disadvantaged group). I.e., $\sigma \to 0$ (more information) maximizes disparities between both groups. Under such conditions, the learner can adopt a consistent policy by revealing partial information, specifically around the region where \emph{Neutrality} occurs for both disparities in scores and utilities.

In the non-monotonic cases where one point of \emph{Neutrality} exists (cases 1 and, 2 with $\mathcal{F}_u(\sigma_{\min}) = 0$; as described in Lemma~\ref{le:bay-ut-cases}), the learner’s optimal strategy closely resembles that of the monotonic decreasing case with a single \emph{Neutrality} point—namely, to identify and operate in a neighborhood around the point of \emph{Neutrality}. Even when no point of \emph{Neutrality} exists in the non-monotonic regime (as seen in Case 2: with $\mathcal{F}_u(\sigma_{\min}) > 0$), the behavior around $\sigma_{\min}$ remains relevant, since it corresponds to the point of minimal \emph{utility disparity}, albeit characterized by \emph{Exploitation}. 

Finally, things become complex when two points of \emph{Neutrality} emerge in the non-monotonic case—i.e., when $\mathcal{F}_u(\sigma_{\min}) < 0$—denoted by $\sigma_{r_1}$ and $\sigma_{r_2}$, with $\sigma_{r_1} < \sigma_{r_2}$. The interval $(\sigma_{r_1}, \sigma_{r_2})$ defines a region of \emph{Burden}, where the learner's intervention results in worsened \emph{utility disparity}. The presence of two \emph{Neutrality} points, $\sigma_{r_1}$ and $\sigma_{r_2}$, poses a significant challenge for the learner. The existence of an intermediate region where \emph{utility disparity} is negative—termed \emph{Burden}—complicates the learner’s signaling strategy, as it is unclear whether revealing more or less information will improve outcomes. In the Lemma below we show a region of selected parameters that corresponds to two distinct points of \emph{Neutrality}.

\begin{lemma}\label{le:region}
In equilibrium, the following is a region $\mathcal{R}$ of relevant parameters that guarantee two points of \emph{Neutrality} for $\mathcal{F}_u$.
\begin{align}
   \mathcal{R} &=\bigg\{k_{\theta_0}, k_{\theta_*}, m, \gamma: k_{\theta_0}^\top k_{\theta_0} < 2  k_{\theta_0}^\top k_{\theta_*} ~\text{and} ~ \gamma > \max\bigg\{\sqrt{\frac{1}{m}\big(2 k_{\theta_0}^\top k_{\theta_*} - k_{\theta_0}^\top k_{\theta_0} + 3 k_{\theta_*}^\top k_{\theta_*}\big) }, \sqrt{\frac{2}{m}\vert k_{\theta_0} - k_{\theta_*}\vert} \bigg\} \bigg\} 
\end{align}
\end{lemma}
Using the above feasibility region, we can infer that, given $k_{\theta_0}^\top k_{\theta_0} < 2 k_{\theta_0}^\top k_{\theta_*}$, a sufficiently large prior variance $\gamma^2$ determined by the model parameters can lead to the existence of two distinct points of \emph{Neutrality} for $\mathcal{F}_u$.


\section{Fairness under Unequal Priors}\label{sec:fair-uneq}

In the previous section, we assumed that agents in both groups shared the same prior distribution. We now move away from this assumption and introduce an additional source of heterogeneity: namely, agents from different groups possess distinct prior means. 

To do so, we adopt and expand on the model of~\citet{bechavod2022information}, where agents are assumed to only have access to a subspace of the overall feature space. In their work, if the true model is given by $\theta_*$, group $1$'s ability to recover $\theta_*$ is limited to a subspace $S_1$ of $\mathbb{R}^d$, and group $2$'s is limited to a subspace $S_2$ of $\mathbb{R}^d$. Similarly, we will assume that for all $g \in \{1,2\}$, group $g$'s ability to reason about the learner's model is limited to subspace $S_g$; however, a significant departure from~\citep{bechavod2022information} is that we still consider \emph{Bayesian} agents\footnote{\citet{bechavod2022information} assumes that $\theta_*$ is perfectly known and taken at face value within $S_g$ for group $g$.}, who may still be uncertain about their knowledge of the model. Formally, we make the following assumption about the agents' priors:



 
\begin{assumption} \label{le:erm-result}
    For all groups $g \in \{1,2\}$, let us denote $\Pi_g$ be the projection operator to subspace $S_g$. We assume that in group $g$, an agent's prior is given by $\pi_g(\theta) \sim \mathcal{N}(\Pi_g \theta_*, \gamma^2 \mathbf{I})$.
\end{assumption}


 \begin{remark}
 This projection to a subspace is especially useful in explaining discrepancies in groups on the basis of their feature characteristics. For instance, in bank loan approvals, suppose group A primarily includes applicants with formal employment---so their features emphasize steady income, tax returns, and employer verification. In contrast, group B may comprise individuals from gig or informal sectors, where key features pertain to inconsistent earnings, alternate credit scores, or cash flows from small businesses. Then each of the groups may only see a small and a different part of the space of features. This can also encode bounded rationality with disparate information about the model, where each group only focuses on the top k features, and may have different understanding of what these top-k features are. See~\cite{bechavod2022information} for more details. 
 \end{remark}

 We note a few assumptions made here for simplicity of exposition. First, we assume that the prior is unbiased within $S_g$, i.e. centered around $\Pi_g \theta_*$. We do so to control for the effect of the alignment of the mean of the prior with the true model; i.e., our disparity results will hold \emph{even in the best case when both groups have a credible prior}. Second, both groups have the same variance term $\gamma^2$--we isolate the effect of differences in prior means solely on disparities.

We also define the amount of overlap between both groups using the metric of~\citet{bechavod2022information}.

\begin{definition}[\citep{bechavod2022information}]
Given a true model $\theta_* \in \mathbb{R}^d$ and projections $\Pi_1, \Pi_2 \in \mathbb{R}^{d \times d}$, the information overlap-proxy between groups $g_1$ and $g_2$ is defined as
\begin{equation*}
    r_{1,2}(\theta_*) := \Vert \Pi_1 \theta_* - \Pi_2 \theta_* \Vert.
\end{equation*}
\end{definition}

We now analyze disparities in the context of the Bayesian model framework established above. Notably, our analysis is agnostic to the specific choice of the true model $\theta_*$ used by the learner; rather, it focuses on the interaction between group-specific feature subspaces (captured by $\Pi_g$) and group-dependent cost matrices (i.e $A_g$). This allows us to study group-level discrepancies that arise purely due to structural differences in feature representations.

\subsection{Score-disparity}\label{sub:uneq-sc}
We first analyze fairness via \textit{score-disparity} similar to that done in Section.~\ref{subsec:fairbay}. All the relevant proofs of the Section can be found in Appendix ~\ref{ap:uneq-sc}

\begin{thm} \label{thm:sc-uneq}
For a Bayesian agent, the \textit{score disparity} between both groups in equilibrium is given by:
\begin{align} \label{eq:sc-bay-uneq}
         \mathcal{F}_s &= {\theta_*}^\top( A_1^{-1} \Pi_1  -  A_2^{-1} \Pi_2){\theta_*} + \big[{\theta_*}^\top (A_1^{-1}  -  A_2^{-1} ){\theta_*} - {\theta_*}^\top( A_1^{-1} \Pi_1  -  A_2^{-1} \Pi_2){\theta_*}\big] \beta_{\gamma}(\sigma).
    \end{align}
\end{thm}

We now characterize the behavior of $\mathcal{F}_s$ and outline the behavior in some natural cases. 

\paragraph{When do Exploitation, Burden, and Neutrality occur?}

\begin{lemma}\label{le:fs-un-exp}
    In equilibrium, Exploitation with respect to $\mathcal{F}_s$ always occurs for all $\sigma \in \mathbb{R}^+$and for all $\theta_* \neq \bar{0}$ if and only if $(A_1^{-1} \Pi_1  -  A_2^{-1} \Pi_2) + ( \Pi_1 A_1^{-1}  -   \Pi_2 A_2^{-1})$ is positive semi-definite. Furthermore, if $\Pi_g A_g^{-1} = A_g^{-1} \Pi_g, ~g \in \{1,2\}$, the condition simplifies to $A_1^{-1} \Pi_1  -  A_2^{-1} \Pi_2$ being positive semi-definite.
\end{lemma}

 \paragraph{Note: }Commutativity of $\Pi_g,  A_g^{-1}$ 
 is satisfied in many natural settings including for instance, when $A_g = \alpha_g \mathbf{I} ~\text{for} ~\alpha_g > 0$ (see Appendix E. of \citet{bechavod2022information}). 

 \begin{lemma}\label{le:nullpi1pi2}
   If $A_1^{-1} \Pi_1  -  A_2^{-1} \Pi_2 \succeq 0$, then $\eta(\Pi_1) \subseteq \eta(\Pi_2)$, where $\eta(B)$ represents the null-space of a matrix $B \in \mathbb{R}^{d \times d}$. If the inequality is strict, then $\Pi_1 = \mathbf{I}$.  
\end{lemma} 

Lemmas~\ref{le:fs-un-exp} and~\ref{le:nullpi1pi2} imply that, under many conditions, when \emph{Exploitation} occurs for all $\sigma \geq 0$ and for every true model $\theta_*$, the feature subspace 
observed by the disadvantaged groups is contained is the feature subspace observed by the advantaged groups\footnote{Note that $\eta(\Pi_1) \subseteq \eta(\Pi_2)$ implies that the induced subspaces satisfy $S_2 \subset S_1$.}. I.e., this corresponds to a situation where the advantaged group always has access to more information compared the disadvantaged group. Furthermore, \emph{Exploitation} persists even in the limit of no information ($\sigma \rightarrow \infty$) for every $\theta_*$ if and only if the advantaged group’s prior fully spans the feature subspace---this is however a corner case that may not arise in realistic scenarios.

We now state our condition for the existence of \emph{Neutrality}:

\begin{lemma}\label{le:fs-un-neut}
    In equilibrium, \textit{Neutrality} with respect to $\mathcal{F}_s$ occurs for all $\theta_* \neq \bar{0}$ if and only if $(A_1^{-1} \Pi_1  -  A_2^{-1} \Pi_2) + (\Pi_1 A_1^{-1}  -   \Pi_2 A_2^{-1})$ is negative definite, at 
    \[
    \sigma = \sigma_r \triangleq \sqrt{ \frac{{\theta_*}^\top (A_1^{-1}  -  A_2^{-1} ){\theta_*}}{{\theta_*}^\top   (\Pi_2 A_2^{-1}  - \Pi_1 A_1^{-1})\theta_*}} \gamma.
    \]
    Furthermore, if $\Pi_g A_g^{-1} = A_g^{-1} \Pi_g, ~g \in \{1,2\}$, then Neutrality occurs for all $\theta_* \neq \bar{0}$ if and only if $A_1^{-1} \Pi_1  -  A_2^{-1} \Pi_2$ is negative definite.
\end{lemma}

\begin{lemma} \label{le:pi1_I_pi2_o}
    Given $A_2 \succ A_1$, if $(A_1^{-1} \Pi_1  -  A_2^{-1} \Pi_2)  \prec 0$, then $\Pi_1 \neq \mathbf{I}$ and  $\Pi_2 = \mathbf{I}$. 
\end{lemma}

Lemmas~\ref{le:fs-un-neut} and~\ref{le:pi1_I_pi2_o} imply that \emph{Neutrality} could occur with respect to $\mathcal{F}_s$ for every true model parameter $\theta_*$ only when the disadvantaged group has full information of the true model $\theta_*$ while the advantaged group does not have full information on $\theta_*$. This is expected to be a rare scenario, in fact showing that we do not expect neutrality to happen in practice with respect to scores for complex, black-box models. In practice, this means that a learner interested in fairness may have to aim for reducing \emph{Exploitation} as much as possible but will never fully achieve \emph{Neutrality}.


\paragraph{How does the level of information revelation affect fairness?} We shift our attention towards monotonicity of $\mathcal{F}_s$ with respect to $\sigma$. 

\begin{definition}
    For an orthogonal projection matrix $\Pi$, its complement $\Pi^\perp$ is defined as $\Pi^\perp = \mathbf{I} - \Pi$.
\end{definition}

We now move on to our main monotonicity result:

\begin{lemma}\label{le:sc-un-dec}
    In equilibrium, $\mathcal{F}_s$ is monotonically non-increasing (respectively, non-decreasing) with $\sigma$ for all $\theta_*$ if and only if $(A_1^{-1} \Pi_1^\perp - A_2^{-1} \Pi_2^\perp) + (\Pi_1^\perp A_1^{-1}  - \Pi_2^\perp A_2^{-1})$
is positive semi-definite (respectively, negative semi-definite). The inequality is strict for all $\theta_* \neq \bar{0}$ — that is, $\mathcal{F}_s$ is strictly decreasing (increasing)—if the matrix is positive definite (negative definite). Furthermore, if $\Pi_g A_g^{-1} = A_g^{-1} \Pi_g$ for each $g \in \{1,2\}$, the condition reduces to $ A_1^{-1} \Pi_1^\perp - A_2^{-1} \Pi_2^\perp$ being positive (negative) semi-definite, or positive (negative) definite for strict monotonicity.
\end{lemma}

The above lemma relies on  conditions: $A_1^{-1} \Pi_1^\perp  -  A_2^{-1} \Pi_2^\perp$ being either positive/negative semi-definite or positive/negative definite. Below, we provide interpretations of these conditions:
\begin{lemma}\label{le:api-perp-spd}
     Given $A_2 \succ A_1$, we have the following:
     \begin{enumerate}
         \item If $(A_1^{-1} \Pi_1^\perp  -  A_2^{-1} \Pi_2^\perp)  \succeq 0$, then $\text{span}(\Pi_1)  \subseteq \text{span}(\Pi_2)$. If the inequality is strict, then $\Pi_1 = \mathbf{0}$.
         \item If $(A_1^{-1} \Pi_1^\perp  -  A_2^{-1} \Pi_2^\perp)  \preceq 0$, then $\text{span}(\Pi_2) \subseteq \text{span}(\Pi_1)$. If the inequality is strict, then $\Pi_1 \neq \mathbf{0}$, and $\Pi_2 = \mathbf{0}$.
     \end{enumerate}  
\end{lemma}

Lemmas~\ref{le:sc-un-dec} and~\ref{le:api-perp-spd} together imply that when $\mathcal{F}_s$ is increasing in $\sigma$ (for all $\theta_*$), the prior feature subspace of the disadvantaged group is contained within that of the advantaged group. In this case, the disadvantaged group is disadvantaged in terms of \emph{both} costs and information. As the learner reveals more information (lowers $\sigma$), he reduces the informational disparities across both groups, helping group $2$; i.e. $\mathcal{F}_s$ becomes smaller at lower $\sigma$'s. Similarly, when $\mathcal{F}_s$ is decreasing in $\sigma$, 
the prior feature subspace of the advantaged group is contained within that of the disadvantaged group. In this case, the learner revealing more information about the model strongly benefits the advantaged group, and disparities increase at low values of $\sigma$. Interestingly, and despite the advantaged group still having an edge in terms of cost of feature manipulations, $\mathcal{F}_s$ remains monotonic.

\paragraph{Overlapping, non-nested subspaces} We now quantify the effect of overlap between prior subspaces of both groups on the disparity in scores at equilibrium. To do so, we let $A_1 = A_2 := A$ to isolate the effect of the information overlap itself, and upper bound $\vert \mathcal{F}_s \vert$ using the \emph{information-overlap proxy} that we defined earlier. 
\begin{lemma}\label{le:sc-proxy}
    Suppose $A_1 = A_2 = A$, then:
    \[\vert \mathcal{F}_s \vert \leq (1 - \beta_{\gamma}(\sigma)) \Vert A\Vert^{-1} \Vert \theta_* \Vert \cdot r_{1,2} ({\theta_*}).
    \]
\end{lemma}
The above inequality relates \emph{score disparity} between two groups to the information-overlap proxy $r_{1,2} ({\theta_*})$ using an upper bound on $\vert \mathcal{F}_s \vert$. The bound decreases as the amount of information increases (i.e the RHS decreases, given all other parameters, as $\sigma$ decreases) and equals $0$ at full information ($\sigma = 0$) whereas at the limit of no information ($\sigma \longrightarrow \infty$ ), the bound goes to $\Vert A\Vert^{-1} \Vert \theta_* \Vert r_{1,2} ({\theta_*})$. The bound suggests that more information overlap, perhaps unsurprisingly, leads to fewer disparities across groups.

\subsection{Utility disparity}\label{sec:ut-uneq}
We now analyze the \emph{utility disparity} for a Bayesian agent when both groups choose their prior means based on the subspace of feature information of model known to them. Throughout the section, we assume commutativity of $A_g^{-1}, \Pi_g$; consistent with the assumptions used in the analysis of the \emph{score disparity}. We start by deriving the \emph{utility disparity} for Bayesian agents:

\begin{thm}\label{thm:ut-uneq}
Let $k_\pi \coloneq  {\theta_*}^\top (A_1^{-1}\Pi_1 - A_2^{-1}\Pi_2) \theta_*$, $k_A\coloneq  {\theta_*}^\top (A_1^{-1} - A_2^{-1}) {\theta_*}$, and $m \coloneq  (\text{Tr}(A_1^{-1}) - \text{Tr}(A_2^{-1}))$. 
If $\Pi_g A_g^{-1} = A_g^{-1} \Pi_g$ for all $g \in \{1,2\}$, the \textit{utility disparity} between both groups in equilibrium is given by: 
\begin{align}\label{eq:ut-uneq}
    \mathcal{F}_u(\sigma) &= \frac{\beta_\gamma^2 (\sigma)}{2} \big(k_\pi - k_A - \sigma^2 m \big  )
    + \beta_\gamma (\sigma)\big( k_A - k_\pi\big)+  \frac{k_\pi}{2}.
\end{align}
\end{thm}

Note that for full and no information revelation, we have:
\[
\mathcal{F}_u (0) = \frac{k_A}{2} > 0, ~ \lim_{\sigma \to \infty} \mathcal{F}_u (\sigma) =  \frac{k_\pi}{2}.
\]
We now characterize all possible behaviors of $\mathcal{F}_u$: 
\begin{lemma}\label{le:bay-ut-uneq}
  In equilibrium, $\mathcal{F}_u$ is bounded. The following are possible behaviors of $\mathcal{F}_u$:
  \begin{enumerate}
      \item \noindent  \textbf{Monotone case: }If $\gamma^2 \leq \frac{2}{m}(k_A - k_\pi)$, then $\mathcal{F}_u$ is monotonically decreasing in $\sigma$. The following are the sub-cases:
      \begin{enumerate}
          \item If $k_\pi < 0$, then $\mathcal{F}_u$ attains \emph{Neutrality} at a unique $\sigma$.
          \item If $k_\pi \geq 0$, then $\mathcal{F}_u$ does not attain \emph{Neutrality}  for all $\sigma$, and instead shows \emph{Exploitation} for all $\sigma$.
      \end{enumerate}
      \item \noindent \textbf{Non-monotone case: } If $\gamma^2 > \frac{2}{m}(k_A - k_\pi)$, then $\mathcal{F}_u$ attains a global minimum at a unique $\sigma_{\min}$ given by:
      \[\sigma_{min} = \sqrt{\frac{1}{1 - \frac{2}{m \gamma^2}(k_A - k_\pi)}}~ \gamma.\]
      $\mathcal{F}_u$ is decreasing for $\sigma \leq \sigma_{\min}$, and increasing for $\sigma > \sigma_{\min}$. The following are sub-cases:
      \begin{enumerate}
          \item If $k_\pi < 0$, then $\mathcal{F}_u$ attains \emph{Neutrality} for some $\sigma < \sigma_{\min}$.
          \item If $k_\pi \geq 0$, then $\mathcal{F}_u$ has no point of \emph{Neutrality} if $\mathcal{F}_u(\sigma_{\min})> 0$, one if $\mathcal{F}_u(\sigma_{\min})= 0$, two if $\mathcal{F}_u(\sigma_{\min})< 0$.
      \end{enumerate}
  \end{enumerate}
\end{lemma}

Similar to Lemma~\ref{le:bay-ut-cases}---the case of equal priors across groups, we observe both monotonic and non-monotonic behavior depending on the value of $\gamma$. In particular, when the value of $\gamma$ is low enough, $\mathcal{F}_u$ is monotone in $\sigma$, and when $\gamma$ is large enough, $\mathcal{F}_u$ exhibit non-monotonicities. However, we note here that the monotonic case may not arise when $k_A < k_\pi$ (unlike in Lemma~\ref{le:bay-ut-cases} where there is always a value of $\gamma$ sporting each regime). Another interesting observation is that: if $k_A\geq k_\pi$, then $\sigma_{\min} \geq \gamma$, otherwise if $k_A<k_\pi$, then $\sigma_{\min} < \gamma$. On the other hand, when both priors are the same, $\sigma_{\min}$ corresponding to the non-monotonic behavior always shows $\sigma_{\min} \geq \gamma$ (from eq.~\ref{eq:sigma_min}).


We now leverage the subspace-based matrix characterizations developed earlier to establish conditions under which $\mathcal{F}_u$ exhibits monotonic and non-monotonic behavior \emph{agnostic} of $\theta_*$, analogous to the approach taken for \emph{score disparities} in the unequal prior setting:

\begin{assumption}\label{as:bay-uneq-ut}
    In what follows, we additionally assume $A_1^{-1} \Pi_1 - A_2^{-1} \Pi_2 \succeq 0$, and either $A_1^{-1} \Pi_1^{\perp} - A_2^{-1} \Pi_2^{\perp} \succeq 0$ or $A_1^{-1} \Pi_1^{\perp} - A_2^{-1} \Pi_2^{\perp} \preceq 0$.
\end{assumption}

The constraint $A_1^{-1} \Pi_1 - A_2^{-1} \Pi_2 \succeq 0$ is equivalent to '\textit{no-burden}' for all $\theta_*$ at the limit of no information ($\sigma \rightarrow \infty$), where as the constraint on the positivity of $A_1^{-1} \Pi_1^{\perp} - A_2^{-1} \Pi_2^{\perp}$ helps control the sign of $k_A - k_\pi$ .
\begin{lemma}\label{le:ut-uneq-extraassum}
    If $A_1^{-1} \Pi_1 - A_2^{-1} \Pi_2 \succeq 0$ and $A_1^{-1} \Pi_1^{\perp} - A_2^{-1} \Pi_2^{\perp} \succeq 0$,  then the following characterization holds for $\mathcal{F}_u$ for all $\theta_*$:
    \begin{enumerate}
        \item \textbf{Monotone case: } If $\gamma^2 \mathbf{I} \preceq \frac{2}{m} (A_1^{-1} \Pi_1^{\perp} - A_2^{-1} \Pi_2^{\perp})$, then $\mathcal{F}_u$ is monotonically decreasing in $\sigma$, and does not attain \emph{Neutrality} at any $\sigma$.
        \item \textbf{Non-monotone case: } If $\gamma^2 \mathbf{I} \succ \frac{2}{m} (A_1^{-1} \Pi_1^{\perp} - A_2^{-1} \Pi_2^{\perp})$, then $\mathcal{F}_u$ attains a global minimum at $\sigma_{\min}$ given by:
        \[\sigma_{min} = \sqrt{\frac{1}{1 - \frac{2}{m \gamma^2}(k_A - k_\pi)}}~ \gamma.\]
        Also, $\mathcal{F}_u$ attains no point of Neutrality if $\mathcal{F}_u (\sigma_{\min}) > 0$, one if $\mathcal{F}_u (\sigma_{\min}) = 0$, two if $\mathcal{F}_u (\sigma_{\min}) < 0$. 
    \end{enumerate}
    
\end{lemma}

    It is important to note that in Lemma~\ref{le:ut-uneq-extraassum}, we expect the monotonic case to not happen in practice simultaneously for all $\theta_*$ as the $\gamma^2 \mathbf{I} \preceq \frac{2}{m} (A_1^{-1} \Pi_1^{\perp} - A_2^{-1} \Pi_2^{\perp})$ requirement implies strict positive definiteness of $A_1^{-1} \Pi_1^{\perp} - A_2^{-1} \Pi_2^{\perp}$ which would here would imply $\Pi_1 = \mathbf{0}$ (as it follows from Lemma~\ref{le:nullpi1pi2}). Using similar arguments, we can show that if we assume $A_1^{-1} \Pi_1^{\perp} - A_2^{-1} \Pi_2^{\perp} \preceq 0$, then monotonic case is unrealistic and non-monotonic case holds for every $\gamma > 0$ and $\theta_*$.The number 
  $\sigma$ where neutrality is attained depends on  the sign of $\sigma_{\min}$ which further depends on $\gamma$. 

We now isolate the effect of information overlap on $\mathcal{F}_u$ by assuming equal cost matrices for both groups:

\begin{lemma}\label{le:ut-overlap}
    Suppose $A_1 = A_2 = A$ and $\Pi_g A_g^{-1} = A_g ^{-1} \Pi_g$ for $g \in \{1,2\}$, then $\vert \mathcal{F}_u \vert \leq \frac{1}{2}(1 - \beta_\gamma(\sigma))^2\Vert A\Vert^{-1} \Vert \theta_*\Vert r_{1,2} ({\theta_*})$. 
\end{lemma}
Similarly to the bound for \emph{score disparity}, this bound decreases as the information increases, i.e., the right-hand side decreases with decreasing $\sigma$. It vanishes under full information ($\sigma = 0$) and approaches $\Vert A \Vert^{-1} \Vert \theta_* \Vert r_{1,2}(\theta_*)$ as $\sigma \to \infty$. However, the key difference is that this bound decays faster due to the presence of $\beta_\gamma^2(\sigma)$, in contrast to the $\beta_\gamma(\sigma)$ factor in the score disparity case.

\section{Conclusion and Limitations}
Our key insight is that theamount of information made available to agents---especially under cost asymmetries---has complex and often counterintuitive implications for fairness. In particular, we show that:
\begin{itemize}
    \item For \emph{naive agents}, transparency reduces utility disparities in a monotonic way, but can sometimes \emph{harm} the \emph{advantaged} group with lower costs by inducing over-investment in uninformative directions. Perhaps surprisingly, information relevation does not impact \emph{expected score disparities}, though it does impact the level of randomness around said score disparities.
    \item For \emph{Bayesian agents}, disparities are \emph{bounded}, and utility disparities are often minimized at intermediate levels of transparency. Counter-intuitively, more transparency is not always better; revealing information leads to a tension between i) helping the advantaged group who, with their lower costs, can take advantage of the additional information more efficiently than the disadvantaged group when it comes to the cost of changing features and ii) additional information helps both groups not invest sub-optimal feature modifications. Effect ii) mostly benefits the higher-cost, disadvantaged group when it comes to efficient feature modifications.
    \item When groups differ in \emph{informational priors}, score and utility disparities are shaped by the \emph{alignment and overlap} of group-specific priors. This expands prior work (e.g., \citet{bechavod2022information}) by introducing Bayesian beliefs, where agents quantify uncertainty around their belief about the deployed classifier.
\end{itemize}

We characterize when \emph{Neutrality}, \emph{Exploitation}, and \emph{Burden} occur, and identify settings where the learner can use the amount of information disclosure as a knob to reduce disparities. We believe that our work advances the understanding of fairness in strategic learning settings. 

\paragraph{Limitations and Future Work.} 
We focus on linear models and Gaussian noise, which makes the analysis tractable and allows to derive useful insights on the level of information a learner should reveal. These models may not capture more complex models and uncertainty, though we believe they are providing useful first-order insights. Further, a learner may be interested not only in fairness, but also on deploying a model that is as accurate as possible; an interesting direction work future work is to characterize \emph{accuracy-fairness} trade-offs as a function of how much information a learner reveals about their scoring or decision rule. We expected that this trade-off will provide further arguments for partial information revelation when agents may try to game the classifier, a phenomenon that is prevalent in practice. 

\section{Acknowledgements}
Juba Ziani's research was supported by NSF CAREER Award IIS-2336236.

\newpage
\bibliographystyle{plainnat}  
\bibliography{references}

\newpage
\appendix
\section{Proofs for the equilibrium computation}\label{ap:eq}

\begin{proof}[\textbf{Proof of Lemma.~\ref{le:delxface}}]

The optimization problem for a Naive agent given by Equation~\ref{eq:opt-naive} is concave for any given realization of $Z$. Therefore, we can solve it using first order conditions as below.
\begin{align*}\label{eq:con-opt}
    \nabla u(\Delta x; g) &= ({\theta_*} + \sigma Z_g) - A_g \Delta x = 0
\end{align*}
where $\Delta x = x' - x$.
The solution to the above equation is given by $\Delta x_g = A^{-1}_g ({\theta_*} + \sigma Z_g)$. The positive-definiteness of $A_g$ guarantees the existence of its inverse.
\end{proof}

\begin{proof}[\textbf{Proof of Lemma.~\ref{le:delxbay}}]
We consider the prior distribution for an agent in group $g$ on ${\theta}$ given by $\pi(\theta) \sim \mathcal{N}(\theta;\omega_g , \gamma^2 \mathbf{I})$, and the likelihood function $\pi_{S | {\theta}}(s) \sim \mathcal{N}(s;\theta_*, \sigma^2 \mathbf{I})    
$ The posterior distribution $\pi^S (\theta)$ is normally distributed with parameters 
 can be computed using the conjugate property of Gaussian priors \citep{murphy2012machine} and is given as 
\begin{align*}
    \theta_s^g &= \omega_g + \beta_\gamma(\sigma) (S_g - \omega_g),\\
    C_s &= \frac{\sigma^2 \gamma^2}{\sigma^2 + \gamma^2}\mathbf{I},
\end{align*}
where $\beta_\gamma(\sigma) \coloneq \frac{\gamma^2}{\gamma^2 + \sigma^2}$. Using the posterior distribution of $\theta$ with the distribution $\pi^S_g (\theta)$, an agent in group $g$ now computes the feature change vector as 
\begin{align*}
    \max_{\Delta x}~~~ \mathbb{E}_{\theta \sim \pi^S_g (\theta)} \left[\theta^\top(x + \Delta x) | S_g \right] - \frac{1}{2}\Delta x^\top A_g \Delta x.
\end{align*}
The solution to this concave program can be found similar to Lemma~\ref{le:delxface} and is given by $\Delta x_g = A_g^{-1} \theta_s^g = A_g^{-1}[\omega_g + \beta_\gamma(\sigma) (S_g - \omega_g)]$.     
\end{proof}

\section{Proofs for the fairness analysis}\label{ap:fairness}
The following are the proofs for propositions in Section~\ref{sec:fair-analysis}.

\subsection{Proofs for Lemmas~\ref{le:scf_naive},~\ref{le:sc-bay}}

\begin{proof}[\textbf{Proof of Lemma.~\ref{le:scf_naive}}]
The expected difference in scores between groups for a \textit{naive} agent is given by:
\begin{align*}
   \mathcal{F}_s &= \mathbb{E} [{\theta_*}^\top \Delta x_1 ] - \mathbb{E} [{\theta_*}^\top \Delta x_2] \nonumber\,\\
    &= \mathbb{E}[{\theta_*}^\top A_1^{-1} ({\theta_*} + \sigma Z_1)] - \mathbb{E}[{\theta_*}^\top A_2^{-1} ({\theta_*} + \sigma Z_2)]\nonumber \,\\
    &= {\theta_*}^\top (A_1^{-1} - A_2^{-1}){\theta_*} \nonumber. 
\end{align*}
    
\end{proof}

\begin{proof}[\textbf{Proof of Corollary.~\ref{co:var-sc}}]
Using that 
\[\Delta x_g = A_g^{-1}(\theta_* + \sigma Z),
\]
we have
    \begin{align*}
    \operatorname{Var}(\theta_*^{\top} \Delta x_1 - \theta_*^{\top} \Delta x_2) 
    & = \sigma^2 \cdot \operatorname{Var}(\theta_*^{\top} A_1^{-1} Z_1 - \theta_*^{\top} A_2^{-1} Z_2) \nonumber\,\\
&= \sigma^2 \theta_*^{\top}(A_1^{-1} A_1^{-\top} + A_2^{-1} A_2^{-\top})\theta_* \nonumber.
    \end{align*} 
\end{proof}

\begin{proof}[\textbf{Proof of Lemma.~\ref{le:utf-naive}}]\label{pr:naiv-ut}
 \begin{align*}
  \mathcal{F}_s  &= \mathbb{E}[{\theta_*}^\top \Delta x_1 - {\theta_*}^\top \Delta x_2] - \mathbb{E}[\frac{1}{2}\Delta x_1^\top A_1 \Delta x_1 - \frac{1}{2}\Delta x_2^\top A_2 \Delta x_2]\nonumber \,\\
    &= {\theta_*}^\top (A_1^{-1} - A_2^{-1}){\theta_*}  - \mathbb{E}[\frac{1}{2}\Delta x_1^\top A_1 \Delta x_1 - \frac{1}{2}\Delta x_2^\top A_2 \Delta x_2].
\end{align*}

Now, we compute $\mathbb{E}[\Delta x_g^\top A_g \Delta x_g]$ as follows:

\begin{align*}
    \mathbb{E}[\Delta x_g^\top  A_g \Delta x_g] 
    &= ({\theta_*} + \sigma Z_g)^\top  A_g^{-1} A_g A_g^{-1} ({\theta_*} + \sigma Z_g) \nonumber\,\\
    &= {\theta_*}^\top A_g^{-1} {\theta_*}  + 2 \sigma {\theta_*}^\top A_g^{-1} Z_g + \sigma^2 Z_g^\top A_g^{-1} Z_g \nonumber\,\\
    &= {\theta_*}^\top A_g^{-1} {\theta_*} + \sigma^2 \text{Tr}(A_g^{-1}).
\end{align*}

Therefore, $\mathcal{F}_u$ is given by:
\begin{align*}
    &{\theta_*}^\top (A_1^{-1} - A_2^{-1}){\theta_*} 
     - \frac{1}{2}\big[{\theta_*}^\top (A_1^{-1} - A_2^{-1}){\theta_*} 
    - (\sigma^2 \text{Tr}(A_1^{-1}) - \sigma^2 \text{Tr}(A_2^{-1}))\big] \nonumber \,\\
    &= \frac{1}{2}{\theta_*}^\top (A_1^{-1} - A_2^{-1}){\theta_*} - \frac{1}{2}(\sigma^2 \text{Tr}(A_1^{-1})- \sigma^2 \text{Tr}(A_2^{-1})).
\end{align*}
   
\end{proof}




\begin{proof}[\textbf{Proof of Lemma.~\ref{co:naiv-ut-beh}}]
We have that $A_2\succ A_1 \implies A_1^{-1} \prec A_2^{-1} \implies Tr(A_1^{-1}) \prec Tr(A_2^{-1})$. Hence $\mathcal{F}_u$ is immediately monotonically decreasing in $\sigma$.
\end{proof}

\subsection{Proofs for Bayesian Agents}

\begin{proof}[\textbf{Proof of Theorem.~\ref{le:sc-bay}}]
We compute the expected difference in scores using $ \Delta x_g= A_g^{-1}  [\omega_g + \beta_\gamma (\sigma)(S_g - \omega_g)]$ (Lemma.~\ref{le:delxbay}). We obtain:
\begin{align*}
    \mathbb{E} [{\theta_*}^\top \Delta x_g ] &= {\theta_*}^\top A_g^{-1}  [\omega_g + \beta_\gamma (\sigma)(\theta_* - \omega_g)] \nonumber\,\\
     &= {\theta_*}^\top A_g^{-1}  \omega_g + \beta_\gamma (\sigma){\theta_*}^\top A_g^{-1} (\theta_* - \omega_g) \nonumber \,\\
     &= {\theta_*}^\top A_g^{-1}  \omega_g + \beta_\gamma (\sigma){\theta_*}^\top A_g^{-1} (\theta_* - \omega_g). \nonumber
\end{align*}
 \begin{align*}\label{eq:sc-bay-deriv}
     &\mathbb{E} [{\theta_*}^\top \Delta x_1 ]  - \mathbb{E} [{\theta_*}^\top \Delta x_2]  = {\theta_*}^\top A_1^{-1} \omega_1 - {\theta_*}^\top A_2^{-1} \omega_2  + \beta_{\gamma}(\sigma)\big[ {\theta_*}^\top (A_1^{-1} - A_2^{-1}){\theta_*} - ({\theta_*}^\top A_1^{-1} \omega_1 - {\theta_*}^\top A_2^{-1} \omega_2) \big].
\end{align*}   
Let $\theta_0 := \omega_1 = \omega_2$, and $k_{\theta_0} = \sqrt{A_1^{-1} - A_2^{-1}}~\theta_0$ and $k_{\theta_*} = \sqrt{A_1^{-1} - A_2^{-1}}~\theta_*$, then the result follows.
\end{proof}

\begin{proof}[\textbf{Proof of Lemma.~\ref{co:sc-bay-prop}}]
First let us prove 1 and 2. We know $\beta_\gamma(\sigma) := \frac{\gamma^2}{\gamma^2 + \sigma^2}$. Then $\mathcal{F}_s$ is monotonically decreasing if the coefficient in front of $\beta_\gamma(\sigma)$ is positive, whereas it is increasing (respectively, non-decreasing) coefficient for $\beta_\gamma(\sigma)$ is negative (not positive). In order to prove 3., we remark that $\beta_\gamma(0) = 1$, and $\lim_{\sigma \to \infty} \beta_\gamma(\sigma) = 0$, and the result follows.
\end{proof}

\begin{proof}[\textbf{Proof of Lemma~\ref{co:bay-sc-neut}}]
From the continuity and monotonicity of $\mathcal{F}_s(\sigma)$, and the fact that $\mathcal{F}_s(0) > 0$ and $\lim_{\sigma \to \infty} \mathcal{F}_s(\sigma) = k_{\theta_0}^\top k_{\theta_*} < 0$ (given), it follows by the Intermediate Value Theorem that there exists a unique $\sigma$ for which $\mathcal{F}_s(\sigma) = 0$, i.e., a point of \emph{Neutrality}.

Conversely, assume a neutrality point exists, but suppose instead that $k_{\theta_0}^\top k_{\theta_*} \geq 0$. Then, by monotonicity, $\mathcal{F}_s(\sigma)$ cannot be less than zero as $\sigma \to \infty$, which contradicts the existence of a \emph{Neutrality}.

We now compute $\sigma_r$ by substituting $\mathcal{F}_s(\sigma_r) = 0$ in eq.~\ref{eq:sc-bay-eq} as follows:
\begin{align*}
    0 &= (1 - \beta_{\gamma}(\sigma_r))\, k_{\theta_0}^\top k_{\theta_*} + \beta_{\gamma}(\sigma_r)\, k_{\theta_*}^\top k_{\theta_*} \\
    &= k_{\theta_0}^\top k_{\theta_*} + \left(k_{\theta_*}^\top k_{\theta_*} - k_{\theta_0}^\top k_{\theta_*} \right)\, \beta_{\gamma}(\sigma_r).
\end{align*}

Rearranging gives:
\begin{align*}
    -k_{\theta_0}^\top k_{\theta_*} &= \left(k_{\theta_*}^\top k_{\theta_*} - k_{\theta_0}^\top k_{\theta_*} \right)\, \frac{\gamma^2}{\gamma^2 + \sigma_r^2}.
\end{align*}

Solving for $\sigma_r^2$:
\begin{align*}
    \frac{\gamma^2}{\gamma^2 + \sigma_r^2} &= \frac{-k_{\theta_0}^\top k_{\theta_*}}{k_{\theta_*}^\top k_{\theta_*} - k_{\theta_0}^\top k_{\theta_*}} \\
    \Rightarrow \gamma^2 + \sigma_r^2 &= \gamma^2 \left( \frac{k_{\theta_*}^\top k_{\theta_*} - k_{\theta_0}^\top k_{\theta_*}}{-k_{\theta_0}^\top k_{\theta_*}} \right) \\
    \Rightarrow \sigma_r &= \sqrt{- \frac{k_{\theta_*}^\top k_{\theta_*}}{k_{\theta_0}^\top k_{\theta_*}}} \gamma.
\end{align*}
\end{proof}

\begin{proof}[\textbf{Proof of Corollary~\ref{co:assum}}]
We use the fact that
  \[\Vert k_{\theta_0} \Vert \leq \Vert k_{\theta_*} \Vert \iff k_{\theta_0}^\top k_{\theta_0}\leq k_{\theta_*}^\top k_{\theta_*}.\] Additionally,
 $k_{\theta_*}^\top k_{\theta_*} + k_{\theta_0}^\top k_{\theta_0} - 2 k_{\theta_0}^\top k_{\theta_*} \geq 0$ (square of a Variable), we have $2 k_{\theta_0}^\top k_{\theta_*} \leq  k_{\theta_*}^\top k_{\theta_*} + k_{\theta_0}^\top k_{\theta_0} \leq 2 k_{\theta_*}^\top k_{\theta_*}$. Therefore $k_{\theta_0}^\top k_{\theta_*} \leq k_{\theta_*}^\top k_{\theta_*}$, and the coefficient of $\beta_\gamma(\sigma)$ is non-negative, hence $\mathcal{F}_s$ is monotonic non-increasing. The upper and lower bounds follow from monotonic non-increasing property.
\end{proof}

\begin{proof}[\textbf{Proof of Theorem~\ref{le:bay-ut-eq}}]
    
We have already computed the expected difference in \emph{scores}. In order to compute the expected difference in \emph{utilities}, we first compute the difference in costs. For a group $g$:
\begin{align*}
    \mathbb{E}[\Delta x_g^\top A_g \Delta x_g]  &= [\omega_g + \beta_\gamma(\sigma) (S_g - \omega_g)]^\top  A_g^{-1} A_g A_g^{-1} [\omega_g+ \beta_\gamma(\sigma) (S_g - \omega_g)]\nonumber \,\\
    &= \omega_g^\top A_g^{-1} \omega_g + 2\beta_\gamma (\sigma)\omega_g^\top A_g^{-1}(S_g - \omega_g)+ \beta_\gamma^2 (\sigma)(S_g - \omega_g)^\top A_g^{-1}(S_g - \omega_g).
\end{align*}
Noting that $S_g = {\theta_*} + \sigma Z_g$, that $\mathbb{E} S_g = {\theta_*}$, and that $\mathbb{E} [Z_g^\top Z_g] = d \sigma^2$:
\begin{align*}
   & \mathbb{E}[(S_g - \omega_g)^\top A_g^{-1}(S_g - \omega_g)] \nonumber \\
   &= \mathbb{E}[S_g^\top A_g^{-1} S_g - 2 S_g^\top A_g^{-1} \omega_g + \omega_g^\top A_g^{-1} \omega_g] \nonumber \\
   &= \mathbb{E}[({\theta_*} + \sigma Z_g)^\top A_g^{-1} ({\theta_*} + \sigma Z_g) - 2 ({\theta_*} + \sigma Z_g)^\top A_g^{-1} \omega_g + \omega_g^\top A_g^{-1} \omega_g] \nonumber \\
   &= \mathbb{E}[{\theta_*}^\top A_g^{-1} {\theta_*} + 2 \sigma {\theta_*}^\top A_g^{-1} Z_g + \sigma^2 Z_g^\top A_g^{-1} Z_g - 2 {\theta_*}^\top A_g^{-1} \omega_g - 2 \sigma Z_g^\top A_g^{-1} \omega_g + \omega_g^\top A_g^{-1} \omega_g] \nonumber \\
   &= {\theta_*}^\top A_g^{-1} {\theta_*} + \sigma^2 \text{Tr}(A_g^{-1}) - 2 {\theta_*}^\top A_g^{-1} \omega_g + \omega_g^\top A_g^{-1} \omega_g.
\end{align*}

\noindent This implies that
\begin{align*}
\mathcal{F}_u 
   & = \mathbb{E}[{\theta_*}^\top \Delta x_1 - {\theta_*}^\top \Delta x_2] - \frac{1}{2}\mathbb{E}[\Delta x_1^\top A_1 \Delta x_1 - \Delta x_2^\top A_2 \Delta x_2]\\
&= {\theta_*}^\top A_1^{-1} \omega_1 - {\theta_*}^\top A_2^{-1} \omega_2 + \beta_{\gamma}(\sigma)\big[ {\theta_*}^\top (A_1^{-1} - A_2^{-1}){\theta_*}- ({\theta_*}^\top A_1^{-1} \omega_1 - {\theta_*}^\top A_2^{-1} \omega_2) \big]  \nonumber \,\\
& - \frac{1}{2}\big[ \omega_1^\top A_1^{-1} \omega_1 -  \omega_2^\top A_2^{-1} \omega_2 
+ 2\beta_\gamma (\sigma)(\omega_1^\top A_1^{-1}{\theta_*} - \omega_2^\top A_2^{-1}{\theta_*} ) - 2\beta_\gamma (\sigma)(\omega_1^\top A_1^{-1} \omega_1 - \omega_2^\top A_2^{-1} \omega_2) \nonumber \,\\
&+ \beta_\gamma^2 (\sigma)({\theta_*}^\top A_1^{-1}{\theta_*} - {\theta_*}^\top A_2^{-1}{\theta_*}) +  \beta_\gamma^2 (\sigma)\sigma^2 (\text{Tr}(A_1^{-1}) - \text{Tr}(A_2^{-1})) - 2\beta_\gamma^2 (\sigma) ({\theta_*}^\top A_1^{-1}\omega_1 - {\theta_*}^\top A_2^{-1}\omega_2) \nonumber \,\\
&+ \beta_\gamma^2 (\sigma)(\omega_1^\top A_1^{-1} \omega_1 - \omega_2^\top A_2^{-1} \omega_2)\big].
\end{align*}

\noindent From now on, to simplify the notations, we let 
\begin{align*}
k_1 &= {\theta_*}^\top A_1^{-1} \omega_1 - {\theta_*}^\top A_2^{-1} \omega_2, \\
k_2 &= \omega_1^\top A_1^{-1} \omega_1 - \omega_2^\top A_2^{-1} \omega_2, \\
k_3 &= {\theta_*}^\top A_1^{-1} {\theta_*} - {\theta_*}^\top A_2^{-1} {\theta_*}.
\end{align*}

The notations $k_1$, $k_2$, and $k_3$ serve as general-purpose notations for constants that arise in the derivation of $\mathcal{F}_u$, corresponding to arbitrarily chosen prior means for both groups. These constants are later transformed and reinterpreted using the specific notation adopted in the main body of the paper, particularly when analyzing the two key cases: (i) a common prior mean shared by both groups, and (ii) distinct prior means constrained to lie within each group's respective feature subspace. We will see this later in the Appendix.

Therefore, the utility-fairness can be represented in condensed version as:
\begin{align*}
  \mathcal{F}_u &= \frac{\beta_\gamma^2 (\sigma)}{2} \big( 2 k_1 - k_2 - k_3 \big) + \beta_\gamma (\sigma)\big( k_2 + k_3 - 2 k_1\big) + \Big( k_1 - \frac{k_2}{2} \Big)  - \frac{1}{2} \beta_\gamma^2 (\sigma)\sigma^2(\text{Tr}(A_1^{-1}) - \text{Tr}(A_2^{-1})).
\end{align*}

\end{proof}
\subsubsection{Analysis of the utility disparity}
We now analyze the behavior of utility function in terms of $\sigma$ by first-order analysis. 
It suffices to analyze the function $\mathcal{F}_u (\sigma)$ given by
\begin{align}\label{eq:ut-fn}
    \mathcal{F}_u (\sigma) &:= \frac{\beta_\gamma^2 (\sigma)}{2} \big( 2 k_1 - k_2 - k_3 \big) + \beta_\gamma (\sigma)\big( k_2 + k_3 - 2 k_1\big)+ \Big( k_1 - \frac{k_2}{2} \Big)  - \frac{1}{2} \beta_\gamma^2 (\sigma)\sigma^2m
\end{align}
where 
\[ k_1 = {\theta_*}^\top A_1^{-1}\omega_1 - {\theta_*}^\top A_2^{-1}\omega_2, \, k_2 = \omega_1^\top A_1^{-1}\omega_1 - \omega_2^\top A_2^{-1}\omega_2, \]  \[k_3 = {\theta_*}^\top A_1^{-1}{\theta_*} - {\theta_*}^\top A_2^{-1}{\theta_*}, m = (\text{Tr}(A_1^{-1}) - \text{Tr}(A_2^{-1}))\]
Computing the first order derivative w.r.t $\sigma$, we have 
\begin{align*}
\mathcal{F}_u ' (\sigma) &= \big( 2 k_1 - k_2 - k_3 \big)\beta(\sigma) \beta'(\sigma)+ \big( k_2 + k_3 - 2 k_1\big) \beta'(\sigma) - m \sigma \beta(\sigma)  (\beta(\sigma) + \sigma \beta'(\sigma) )
\end{align*}
\[\text{Using}\,\, \beta'(\sigma) = \frac{-2 \gamma^2 \sigma}{(\gamma^2 + \sigma^2)^2} \,\, \text{and substituting}\,\, \beta, \beta' ~\text{into} ~\mathcal{F}_u ' (\sigma)\]
We now have
\begin{align*}
&\mathcal{F}_u ' (\sigma) = \big( 2 k_1 - k_2 - k_3 \big) \frac{\gamma^2}{\gamma^2 + \sigma^2}  \frac{-2 \gamma^2 \sigma}{(\gamma^2 + \sigma^2)^2}+ \big( k_2 + k_3 - 2 k_1\big) \frac{-2 \gamma^2 \sigma}{(\gamma^2 + \sigma^2)^2} - m \sigma \Big( \frac{\gamma^2}{\gamma^2 + \sigma^2}\Big)^2 \nonumber \,\\
&- m \sigma^2 \Big( \frac{\gamma^2}{\gamma^2 + \sigma^2}\Big)  \frac{-2 \gamma^2 \sigma}{(\gamma^2 + \sigma^2)^2}\nonumber \,\\
&= \frac{\sigma \gamma^2}{(\gamma^2 + \sigma^2)^3} \big[-2( 2 k_1 - k_2 - k_3 \big) \gamma^2 - 2 \big( k_2 + k_3 - 2 k_1\big) \gamma^2- 2 \big( k_2 + k_3 - 2 k_1\big) \sigma^2 - m \gamma^4 - m \gamma^2 \sigma^2 + 2m \gamma^2 \sigma^2 \big] \nonumber \,\\
&=  \frac{\sigma \gamma^2}{(\gamma^2 + \sigma^2)^3} \big[ - 2 \big( k_2 + k_3 - 2 k_1\big) \sigma^2 - m \gamma^4 + m \gamma^2 \sigma^2 \big] \nonumber \,\\
&=  \frac{\sigma \gamma^2}{(\gamma^2 + \sigma^2)^3}\big[(m \gamma^2 - 2 ( k_2 + k_3 - 2 k_1)) \sigma^2 - m \gamma^4 \big]. 
\end{align*}

Computing values of $\mathcal{F}_u(\sigma)$ at boundaries, we have: \[ \mathcal{F}_u (0) = \frac{k_3}{2}, \, \lim_{\sigma \to \infty} \mathcal{F}_u (\sigma) = k_1 - \frac{k_2}{2}.
\]

We now list all possible cases of behavior of $\mathcal{F}_u$ from its first order analysis as below:
\begin{itemize}
    \item If $2 ( k_2 + k_3 - 2 k_1) \geq m \gamma^2$, then $\mathcal{F}_u ' (\sigma) < 0$ for all  $\sigma \geq 0$ and therefore $\mathcal{F}_u$ is monotonically decreasing with $\sigma$.
    \item If $2 ( k_2 + k_3 - 2 k_1) < m \gamma^2$, then $\exists\,\, \sigma_{min} > 0: \mathcal{F}_u'(\sigma_{\min}) = 0$ and $\mathcal{F}_u$ attains a global minimum at $\sigma_{\min}$, where $\sigma_{\min}$ is given by:
    \begin{align*}
        \sigma_{\min} &= \sqrt{\frac{m \gamma^4}{m \gamma^2 - 2 ( k_2 + k_3 - 2 k_1)}}~\gamma.
    \end{align*}
\end{itemize}
An important part of our analysis is to find out points for which \emph{Neutrality} happens for both groups, which here means to look for roots of $\mathcal{F}_u (\sigma)$.
Using the positive-definiteness of $A_1^{-1} - A_2^{-1}$, we have $\mathcal{F}_u (0) = \frac{k_3}{2} > 0$. Now the asymptotic behavior $\lim_{\sigma \to \infty}, \mathcal{F}_u (\sigma) = k_1 - \frac{k_2}{2}$ whether positive or negative determines the number of zeros (as a consequence of monotonicity and the Intermediate Value Theorem). We now provide an exhaustive list of all possible cases:

\subsubsection*{Case 1: $\mathcal{F}_u$ is monotonic decreasing}
Assuming $  (k_2 + k_3 - 2 k_1) \geq \frac{m \gamma^2}{2}$, the following are sub-cases:
\begin{itemize}
    \item If $k_2 > 2 k_1$, then $\mathcal{F}_u$ has \emph{Neutrality} at \emph{one} point.
    \item If $k_2 \leq 2 k_1$, then \emph{Neutrality} occurs at \emph{no} point.
\end{itemize}
\subsubsection*{Case 2: $\mathcal{F}_u$ is non-monotonic with one global minimum}
Assuming $  (k_2 + k_3 - 2 k_1) < \frac{m \gamma^2}{2}$, the following are sub-cases:
\begin{itemize}
    \item If $k_2 > 2 k_1$, then $\mathcal{F}_u$ attains \emph{one} point of \emph{Neutrality} for some $\sigma < \sigma_{\min}$.
    \item If $k_2 \leq 2 k_1$, then $\mathcal{F}_u$ has a) two \emph{Neutrality} points if $f(\sigma_{\min}) < 0$ b) one \emph{Neutrality} point if $f(\sigma_{\min}) = 0$,  c) Zero \emph{Neutrality} points otherwise i.e if $f(\sigma_{\min}) > 0$.
\end{itemize}



Using the analysis performed so far for $\mathcal{F}_u$ for arbitrary prior means $\omega_1, \omega_2$, we now derive results for specific cases of $\omega_1, \omega_2$ taken in this work.

\begin{proof}[\textbf{Proof of Theorems~\ref{le:bay-ut-eq} and~\ref{le:bay-ut-cases}}]
Let $\omega_1=\omega_2=\theta_0$. Introduce the Variables $k_{\theta_0} = \sqrt{A_1^{-1} - A_2^{-1}}~\theta_0$ and $k_{\theta_*} = \sqrt{A_1^{-1} - A_2^{-1}}~\theta_*$. We have $k_1 = k_{\theta_0}^\top k_{\theta_*},  k_2 = k_{\theta_0}^\top k_{\theta_0}, k_3 = k_{\theta_*}^\top k_{\theta_*}$.  

Therefore $k_2 + k_3 - 2 k_1 = k_{\theta_*}^\top k_{\theta_*} + k_{\theta_0}^\top k_{\theta_0} - 2 k_{\theta_0}^\top k_{\theta_*} = (k_{\theta_0} - k_{\theta_*})^2$.
Substituting the values of $k_1, k_2, k_3$ in terms of Variables $k_{\theta_*}, k_{\theta_0}$ into the expression for $\mathcal{F}_u(\sigma)$, we prove Theorem~\ref{le:bay-ut-eq}. Now define, 
\[
\gamma_c \coloneq \sqrt{\frac{2}{m}} ~ \vert k_{\theta_*} - k_{\theta_0}\vert.
\]
Substituting the above expression into the term $k_2 + k_3 - 2 k_1$ in \textbf{Case 1}, followed by \textbf{Case 2} of the analysis of $\mathcal{F}_u$, we obtain the form shown in Theorem~\ref{le:bay-ut-cases}.
\end{proof}

\begin{proof}[\textbf{Proof of Lemma~\ref{le:region}}]

From Lemma~\ref{le:bay-ut-cases}, consider Case 2 corresponding to the \textit{non-monotonic} regime in the analysis of $\mathcal{F}_u$. In this case, the following condition must be satisfied:
     \[
     k_{\theta_0}^\top k_{\theta_0} \leq 2 k_{\theta_0}^\top k_{\theta_*}.
     \]

     Now, from Equation~\ref{eq:sigma_min}, using the fact that \[\sigma_{\min}  = \sqrt{\frac{1}{1 - \frac{\gamma_c^2}{ \gamma^2}}}~ \gamma, ~ \gamma_c \coloneq \sqrt{\frac{2}{m}} ~ \vert k_{\theta_*} - k_{\theta_0}\vert; \]
  along with the following observations: a) $ \sigma_{\min} \geq \gamma$, and b) $\mathcal{F}_u$ is decreasing up to $\sigma_{\min}$; we can choose $\gamma$ large enough such that $\mathcal{F}_u (\sigma_{\min}) < 0$. Thus, we have:
   \[
   \mathcal{F}_u (\gamma) =  \frac{2 k_{\theta_0}^\top k_{\theta_*} - k_{\theta_0}^\top k_{\theta_0} + 3 k_{\theta_*}^\top k_{\theta_*}}{8} - \frac{m \gamma^2}{8}.
   \]
 $\sigma_{\min} \geq \gamma \implies \mathcal{F}_u(\sigma_{\min}) \leq \mathcal{F}_u(\gamma)$. 
 Enforcing $\mathcal{F}_u(\gamma) < 0$ by choosing a large enough $\gamma$, we have the condition
 \[\gamma >  \sqrt{\frac{1}{m}\big(2 k_{\theta_0}^\top k_{\theta_*} - k_{\theta_0}^\top k_{\theta_0} + 3 k_{\theta_*}^\top k_{\theta_*}\big) }.\]
By intersecting the above inequality with the condition $\gamma > \gamma_c$, then taking the maximum over $\gamma$ on the right-hand side, and finally intersecting with the initial condition $k_{\theta_0}^\top k_{\theta_0} \leq 2 k_{\theta_0}^\top k_{\theta_*}$, we obtain the desired result.

\end{proof}
\section{Proofs for Unequal Priors}

\subsection{Proofs for Section~\ref{sub:uneq-sc}}\label{ap:uneq-sc}
\begin{proof}[\textbf{Proof of Theorem~\ref{thm:sc-uneq}}: ]
    In eq.~\ref{eq:ut-fn}, let $\omega_1 = \Pi_1 \theta_*$, $\omega_2 = \Pi_2 \theta_*$, we get the required result.
\end{proof}
\begin{proof}[\textbf{Proof of Lemma~\ref{le:fs-un-exp}: } ]
\label{pr:pd-exp} Let $\Omega := A_1^{-1} - A_2^{-1}$ and $M :=  A_1^{-1}\Pi_1 - A_2^{-1}\Pi_2$. Given the condition $\mathcal{F}_s > 0$ (by definition of \textit{Exploitation}), we have:
\begin{align*}
    {\theta_*}^\top M \theta_* + ({\theta_*}^\top \Omega \theta_* - {\theta_*}^\top M \theta_*) \beta_\gamma(\sigma) > 0 ~\forall \sigma\geq 0.
\end{align*}
Monotonicity of $\beta_\gamma (\sigma)$ with respect to $\sigma$ implies $\mathcal{F}_s$ is monotonic in $\sigma$. From continuity of $\mathcal{F}_s$, it is enough to show $\mathcal{F}_s > 0$ at the domain boundary $\sigma = 0$, i.e., $\mathcal{F}_s (0) \equiv {\theta_*}^\top \Omega \theta_* >0$ and that $\mathcal{F}_s$ asymptotically converges to a non-negative value, i.e., $\lim_{\sigma \to \infty}  \mathcal{F}_s \equiv {\theta_*}^\top M \theta_* \geq 0$. ${\theta_*}^\top \Omega \theta_* > 0$ from positive-definiteness of $\Omega$. The only required condition is ${\theta_*}^\top M \theta_* > 0 ~\forall \theta_* \neq \bar{0}$. From standard linear algebra arguments, ${\theta_*}^\top M \theta_* = \frac{1}{2}{\theta_*}^\top (M + M^\top) \theta_*$, Therefore ${\theta_*}^\top M \theta_* > 0 \iff {\theta_*}^\top (M + M^\top) \theta_* > 0$, Since $(M + M^\top)$ is symmetric, we require $M + M^\top \succ 0$. The extension to symmetric $M$ and the relaxation of inequality to positive semi-definiteness is straightforward.
\end{proof}


\begin{proof}[\textbf{Proof of Lemma~\ref{le:nullpi1pi2}}]
 Let $y \neq \bar{0} \in \eta(\Pi_1)$  where $\eta(.)$ is the the null space of a square matrix. Assume on the contrary that $y \notin \eta(\Pi_2)$. Then $y^\top(A_1^{-1} \Pi_1  -  A_2^{-1} \Pi_2)y = - (\Pi_2 y)^\top A_2^{-1} (\Pi_2 y) < 0$ (since $\Pi_2 y \neq \bar{0}$ by definition) which contradicts positive semi-definiteness of $A_1^{-1} \Pi_1  -  A_2^{-1} \Pi_2$. For second part of the lemma, assume on the contrary $\Pi_1 \neq \mathbf{I}$, now choose $y \in \eta(\Pi_1)$. We can contradict positive-definiteness of the objective using similar arguments.   
\end{proof}


\begin{proof}[\textbf{Proof of Lemma~\ref{le:fs-un-neut}}]
Using similar arguments as in the Proof of Lemma~\ref{co:bay-sc-neut}, we have the following: for \emph{Neutrality} to exist, $\theta_*^\top (A_1^{-1}\Pi_1 - A_2^{-1}\Pi_2)\theta_* < 0$ should hold for all $\theta_* \neq \bar{0}$. Let $M:= A_1^{-1}\Pi_1 - A_2^{-1}\Pi_2$. From using similar arguments as previously done in Proof of Lemma~\ref{le:fs-un-exp}, $\theta_*^\top M \theta_* < 0 ~ \forall \theta_* \neq \bar{0} \iff (M + M^\top) \prec 0$. To find $\sigma_r$, we solve for $\mathcal{F}_s(\sigma_r) = 0$ whose procedure is very similar to that done in Proof of Lemma~\ref{co:bay-sc-neut}.
\end{proof}

\begin{proof}[\textbf{Proof of Lemma~\ref{le:pi1_I_pi2_o}}]
 On the contrary assume $\Pi_2 \neq \mathbf{I}$. Then $\Pi_2$ is not full rank (as the only full rank orthogonal projector is Identity). Therefore, $\exists y \neq \bar{0} \in \mathbb
R^d : \Pi_2 y = \bar{0}$. Then $y^\top(A_1^{-1} \Pi_1  -  A_2^{-1} \Pi_2)y = (\Pi_1 y)^\top A_1 (\Pi_1 y) \geq 0$ which contradicts the negative-definiteness of $A_1^{-1} \Pi_1  -  A_2^{-1} \Pi_2$. Hence $\Pi_2 = \mathbf{I}$. Now let $\Pi_1 = \mathbf{I}$, then $A_1^{-1} \Pi_1  -  A_2^{-1} \Pi_2 = A_1^{-1}   -  A_2^{-1}$ which is positive-definite, a contradiction. Hence $\Pi_1 \neq \mathbf{I}$.
\end{proof}

\begin{proof}[\textbf{Proof of Lemma~\ref{le:sc-un-dec}}]
We prove the result for the case where $\mathcal{F}_s$ is \emph{non-increasing}, noting that the extension to the strict inequality case is straightforward.

Recall from Eq.~\ref{eq:sc-bay-uneq} that $\mathcal{F}_s$ is non-increasing if and only if the coefficient of $\beta_\gamma(\sigma)$ is non-negative, i.e.,
 \[{\theta_*}^\top (A_1^{-1}  -  A_2^{-1} ){\theta_*} 
    - {\theta_*}^\top( A_1^{-1} \Pi_1  -  A_2^{-1} \Pi_2){\theta_*} \geq 0.
\] 
Rewriting, we obtain:
\[    {\theta_*}^\top(A_1^{-1} \Pi_1^\perp - A_2^{-1} \Pi_2^\perp){\theta_*} \geq 0. 
    \quad \]
This inequality holds if and only if 
\begin{equation}
    (A_1^{-1} \Pi_1^\perp - A_2^{-1} \Pi_2^\perp) 
    + (\Pi_1^\perp A_1^{-1} - \Pi_2^\perp A_2^{-1}) \succeq 0.
\end{equation}
The converse follows directly, and the result readily extends to the case of \emph{strict inequality} by replacing $\succeq$ with $\succ$ in the final condition.
\end{proof}

\begin{proof}[\textbf{Proof of Lemma~\ref{le:api-perp-spd}}]
Part 1 can be proven 
in a similar manner as  proof to  Lemma~\ref{le:nullpi1pi2}.  Additionally, using $\text{span}(\Pi) = \eta(\Pi^{\perp})$, we get the required result. To prove that $\Pi_1 \neq \mathbf{0}$, and $\Pi_2 = \mathbf{0}$ if the inequality is strict, we can use the strict inequality result from Lemma~\ref{le:nullpi1pi2} and then use  $\text{span}(\Pi) = \eta(\Pi^{\perp})$. Part 2 follows the same procedure as Part 1. This proves the result.  
\end{proof}

\begin{proof}[\textbf{Proof of Lemma~\ref{le:sc-proxy}}]
Let $A_1 = A_2 \coloneq A$ in eq.~\ref{eq:sc-bay-uneq}, we have:
\begin{align}
         \mathcal{F}_s &= (1 - \beta_\gamma(\sigma)){\theta_*}^\top( A_1^{-1} \Pi_1  -  A_2^{-1} \Pi_2){\theta_*}. \nonumber 
    \end{align}
Taking the 2-norm on both sides of the equation, followed by using the sub-multiplicative property of matrix norms, we get the desired result.
\end{proof}

\subsection{Proofs for Section~\ref{sec:ut-uneq}}
\begin{proof}[\textbf{Proof of Theorem~\ref{thm:ut-uneq}, Lemma~\ref{le:bay-ut-uneq}}]

From eq.~\ref{eq:ut-fn}:
\begin{align*}
    \mathcal{F}_u (\sigma) &:= \frac{\beta_\gamma^2 (\sigma)}{2} \big( 2 k_1 - k_2 - k_3 \big) + \beta_\gamma (\sigma)\big( k_2 + k_3 - 2 k_1\big)+ \Big( k_1 - \frac{k_2}{2} \Big)  - \frac{1}{2} \beta_\gamma^2 (\sigma)\sigma^2m.
\end{align*}
The commutativity assumption $\Pi_g A_g^{-1} = A_{g}^{-1} \Pi_g$ implies that for the equation derived above,
given
\[ k_1 = {\theta_*}^\top A_1^{-1}\omega_1 - {\theta_*}^\top A_2^{-1}\omega_2, \, k_2 = \omega_1^\top A_1^{-1}\omega_1 - \omega_2^\top A_2^{-1}\omega_2, \]  \[k_3 = {\theta_*}^\top A_1^{-1}{\theta_*} - {\theta_*}^\top A_2^{-1}{\theta_*}, m = (tr(A_1^{-1}) - tr(A_2^{-1}));\]
using $\omega_1 = \Pi_1 \theta_*$, $\omega_2 = \Pi_2 \theta_*$, we get
\begin{align*}
    k_1 &= \theta_*^\top(A_1^{-1}\Pi_1 - A_2^{-1} \Pi_2)\theta_* \nonumber\\ &= \theta_*^\top(A_1^{-1}\Pi_1^2 - A_2^{-1} \Pi_2^2)\theta_* \nonumber\\
    &= \theta_*^\top(\Pi_1 A_1^{-1}\Pi_1 - \Pi_2 A_2^{-1} \Pi_2)\theta_* \nonumber\\
    &= k_2. \nonumber
\end{align*}

Note that in the above equations, we used the property of projection matrices i.e $\Pi_g^2 = \Pi_g$, and the commutativity assumption, i.e $\Pi_g A_g^{-1} = A_{g}^{-1} \Pi_g$. Now, define $k_\pi \coloneq k_1 = k_2$, and $k_A \coloneq k_3$ and replacing $k_1, k_2, k_3$ in eq.~\ref{eq:ut-fn} with $k_\pi, k_A$; we get the necessary equation with the relevant Variables.
\end{proof}

\begin{proof}[\textbf{Proof of Lemma~\ref{le:ut-uneq-extraassum}}]
We know
\[k_{\pi} = \theta_*^\top(A_1^{-1}\Pi_1 - A_2^{-1} \Pi_2)\theta_*, ~ k_A =\theta_*^\top(A_1^{-1} - A_2^{-1} )\theta_*.\] 
First, observe that
\[\mathcal{F}_u (0) = \frac{k_A}{2} > 0, ~ \lim_{\sigma \to \infty} \mathcal{F}_u (\sigma) =  k_\pi -\frac{k_\pi}{2} = 
\frac{k_\pi}{2}.\]
Therefore, the first assumption in Assumption.~\ref{as:bay-uneq-ut} leads to $\lim_{\sigma \to \infty} \mathcal{F}_u (\sigma) =  k_\pi -\frac{k_\pi}{2} = 
\frac{k_\pi}{2} \geq 0$.

From Lemma ~\ref{le:bay-ut-uneq}, the following are the conditions that need to be satisfied by $\gamma$ to lie in the following regimes:\\

\noindent \textbf{Monotonic case: } For $\mathcal{F}_u$ to be monotonic, $\gamma^2 \leq \frac{2}{m}(k_A - k_\pi)$. Using the fact that
\begin{align*}
  k_A - k_\pi &= {\theta_*}^\top (A_1^{-1}  -  A_2^{-1} ){\theta_*} - {\theta_*}^\top( A_1^{-1} \Pi_1  -  A_2^{-1} \Pi_2){\theta_*}  \nonumber \,\\
  &= {\theta_*}^\top( A_1^{-1} \Pi_1^\perp  -  A_2^{-1} \Pi_2^\perp){\theta_*},
\end{align*}
and using the positive semi-definiteness of R.H.S, the condition for monotonicity to be satisfied for all $\theta_*$ becomes
\[\gamma^2 \mathbf{I} \preceq \frac{2}{m}(A_1^{-1} \Pi_1^\perp  -  A_2^{-1} \Pi_2^\perp).\]
From Case 1b. of Lemma~\ref{le:bay-ut-uneq}, $\mathcal{F}_u$ does not have any \emph{Neutrality} point and shows \emph{Exploitation} for all $\sigma$. \\

\noindent \textbf{Non-monotonic case: }  For $\mathcal{F}_u$ to be non-monotonic, $\gamma^2 > \frac{2}{m}(k_A - k_\pi)$. Similar to the above Monotonic case, we get 
\[\gamma^2 \mathbf{I} \succ \frac{2}{m}(A_1^{-1} \Pi_1^\perp  -  A_2^{-1} \Pi_2^\perp).\]
From Case 2b, the number of unique points where \emph{Neutrality} is attained depends on the sign of $\sigma_{\min}$.
\end{proof}

\begin{proof}[\textbf{Proof of Lemma~\ref{le:ut-overlap}}]
Substituting $A_1 = A_2 \coloneq A$ in eq.~\ref{eq:ut-uneq} given below
\begin{align*}
  \mathcal{F}_u(\sigma) &= \frac{\beta_\gamma^2 (\sigma)}{2} \big(k_\pi - k_A - \sigma^2 m \big  )
  + \beta_\gamma (\sigma)\big( k_A - k_\pi\big)  +  \frac{k_\pi}{2}, \nonumber
\end{align*}
we get, 
\begin{align*}
     \mathcal{F}_u(\sigma) &= \frac{\beta_\gamma^2 (\sigma)}{2} \big(k_\pi \big  )
    + \beta_\gamma (\sigma)\big(- k_\pi\big) +  \frac{k_\pi}{2}  \nonumber \,\\
    &= \frac{1}{2}(1 - \beta_\gamma(\sigma))^2k_\pi. \nonumber
\end{align*}
Taking the 2-norm on both sides and using the sub-multiplicative property and the definition of \emph{information-overlap proxy}, we have
\begin{align*}
\vert \mathcal{F}_u \vert \leq \frac{1}{2}(1 - \beta_\gamma(\sigma))^2\Vert A\Vert^{-1} \Vert \theta_*\Vert r_{1,2} ({\theta_*}).\nonumber
\end{align*}
\end{proof}

\end{document}